**Highly-Sensitive Resonance-Enhanced Organic Photodetectors for Shortwave Infrared Sensing**

*Hoang Mai Luong[1,2], Chokchai Kaiyasuan[1,3], Ahra Yi[1,4], Sangmin Chae[1,2], Brian Minki Kim[1], Patchareepond Panoy[1,3], Hyo Jung Kim[4], Vinich Promarak[3], Yasuo Miyata[5], Hidenori Nakayama[5], and Thuc-Quyen Nguyen[1,2]\**

[1] Center for Polymers and Organic Solids, University of California, Santa Barbara, CA 93106, USA
\* E-mail: quyen@chem.ucsb.edu

[2] Mitsubishi Chemical Center for Advanced Materials, Materials Research Laboratory, University of California, Santa Barbara, CA 93106, USA

[3] Department of Materials Science and Engineering, School of Molecular Science and Engineering, Vidyasirimedhi Institute of Science and Technology, Wangchan, Rayong 21210, Thailand

[4] Department of Organic Material Science and Engineering, School of Chemical Engineering, Pusan National University, Busan 46241, Republic of Korea

[5] Organic Electronics Group, Information & Electronics Technology Center, R&D Division, Specialty Materials Business Group, Mitsubishi Chemical Corporation, 1000 Kamoshida-cho, Aoba-ku, Yokohama, Kanagawa 227-8502, Japan

**Abstract**

Shortwave infrared (SWIR) has various applications, including night vision, remote sensing, and medical imaging. SWIR organic photodetectors (OPDs) offer advantages such as flexibility, cost-effectiveness, and tunable properties, however, lower sensitivity and limited spectral coverage compared to inorganic counterparts are major drawbacks. Here, we propose a simple yet effective and widely applicable strategy to extend the wavelength detection range of OPD to a longer wavelength, using resonant optical microcavity. We demonstrate a proof-of-concept in PTB7-Th:COTIC-4F blend system, achieving external quantum efficiency (EQE) > 50 % over a broad spectrum $\lambda$ = 450 – 1100 nm with a peak specific detectivity ($D^*$) of $1.1 \times 10^{13}$ Jones at $\lambda$ = 1100 nm, while cut-off bandwidth, speed, and linearity are preserved. By employing a novel small-molecule acceptor IR6, a record high EQE = 35 % and $D^* = 4.1 \times 10^{12}$ Jones are obtained at $\lambda$ = 1150 nm. This research emphasizes the importance of optical design in optoelectronic devices, presenting a considerably simpler method to expand the photodetection range compared to a traditional approach that involves developing absorbers with narrow optical gaps.

## 1. Introduction

Shortwave infrared (SWIR, usually corresponds to the wavelength range of $\lambda = 1000 - 2500$ nm) and their photodetection platforms have attracted elevated attention in recent years.[1-2] SWIR covers two biological windows (1000 – 1350 nm and 1500 – 1800 nm) with low-absorption of biological tissues and fluids,[3-4] has excellent penetrating ability through dusty/foggy air-condition, and enhances visibility in nighttime compared to visible wavelength range.[1, 5] These characteristics give this wavelength range ideal applicability for emerging technologies such as autonomous vehicles, medical devices, industrial inspection, telecommunications, remote sensing, and environmental monitoring.[6-7] Photodetectors for SWIR are mostly based on inorganic materials, notably InGaAs, Ge, and InSb, owing to their low exciton binding energy, high absorption coefficient, high charge mobility, great reproducibility, and stability.[8] Nevertheless, these already-commercialized platforms usually require molecular beam epitaxy to grow the light-absorbing active layer, and the high-production cost is often regarded as the biggest disadvantage.[1, 9]

Solution-processible organic photodetectors (OPDs) are arising as a cost-effective near-infrared (NIR) and infrared (IR) sensing platform, along with other attractive properties such as lightweight, flexible, compatible with mass-production processes (roll-to-roll, blade coating, etc.).[10-16] Recent developments of non-fullerene acceptors (NFAs) and extensive efforts on device engineering have enabled high-detectivity NIR OPDs (~$10^{13}$ Jones at $\lambda = 940$ nm),[17-20] with impressive external quantum efficiency (EQE > 50 % at $\lambda = 940$ nm),[19-23] low dark current density at a reverse bias (sub-nA/cm$^2$),[12, 24-28] and mega-hertz cut-off bandwidth.[29-31] Remarkably, some NFA-based OPDs have demonstrated sizable photoresponse in the SWIR range of 1000 nm $\leq \lambda <$ 1100 nm.[17-18, 21-22, 32-35] For instance, by adopting an ultrathick active layer (~8.2 µm) of PD004:PD-A2 bulk heterojunction (BHJ), Tsai *et al.* realized a self-filtered OPD with EQE = 53 %, at −8 V and $\lambda$ = 1080 nm.[35] Yet, to the best of our knowledge, there are no high-performance NFA-based OPD with EQE > 20 % for $\lambda \geq$ 1100 nm. Beyond this wavelength, SWIR absorption of OPD in the literature majorly comes from (**i**) ultra-narrow bandgap polymer[36-42] or (**ii**) resonance-enhanced absorption of donor:acceptor charge-transfer (CT) state using cavity structure,[30, 43-46] and each of them has their own drawbacks. SWIR OPDs based on (**i**) suffer from low EQE (~10 % or lower) due to poor dissociation coefficient, and therefore charge extraction process heavily depends on the reverse bias.[40] SWIR OPDs based on (**ii**) only support narrow-band detection spectrum, and poor EQE (< 10 %) and low detectivity are generally observed.[46-47] Summarily, extending the OPD detection range to a longer SWIR wavelength while preserving high photoresponse, speed, and broadband detection spectra, is a very challenging task regardless of the mechanisms and materials used.

In this work, we establish a simple yet effective and elegant strategy to extend the wavelength detection range of OPD to a longer wavelength, using resonant optical microcavity. By matching the thickness of the active layer to the optical resonant conditions, we obtain resonant-cavity enhanced (RCE) EQE at the long-wavelength absorption tail of the BHJ blend. This RCE concept in a glass/ITO/ZnO/BHJ/MoO$_x$/Ag OPD is demonstrated using finite-difference time domain (FDTD) simulation. RCE OPDs using PTB7-Th:COTIC-4F BHJ showcase an excellent experimental EQE of 59 % at $\lambda$ = 1100 nm (~5-fold enhancement compared to off-resonant condition) and 34 % at $\lambda$ = 1130 nm (~8-fold enhancement), respectively, while broadband wavelength spectra, bandwidth, speed, and linearity are preserved. Importantly, this work also underlines a possible source of artifact in the analysis using the tail of EQE spectra (*e.g.,* determination of CT-state, mid-gap trap state, Urbach energy fitting, etc.), which originates from resonant-enhanced EQE.

## 2. Results and Discussion

### 2.1. Concept of resonant-cavity enhanced (RCE) OPDs

The architecture of an RCE OPD is simple. The BHJ is sandwiched between a highly-transparent bottom contact (glass/ITO/ZnO, acts as a "front mirror") and an opaque top contact (MoO$_x$/Ag, acts as a "back mirror") (**Figure S1**), and this OPD configuration is a regular bottom-illuminated inverted structure. We note that the "front mirror" semitransparent electrodes of other resonance-enhanced OPDs in the literature are usually formed by a thin layer (~20-30 nm) of Au or Ag, which is the major structural difference compared to our RCE OPD.[30, 43] Two major advantages achieved from this unique highly-transparent electrode are (1) preserving broadband detection spectra (see discussion of **Supplementary Figures S2-3**) and (2) reduced parasitic absorption loss due to the bottom electrode. Similar to any RCE photodetector, the resonance condition of the RCE OPD can be estimated from the following equation,[30, 43, 48]

$$m = \frac{2n_{eff}L}{\lambda_m}, \qquad (1)$$

where $\lambda_m$, $n_{eff}$, $L$, $m$ are resonance wavelength, effective refractive index of the medium between two mirrors, thickness of the medium between two mirrors, and resonance mode order, respectively.

**Figure 1a** shows how we use this RCE concept for device design. Taking an example of PTB7-Th:COTIC-4F blend which has been featured in several recent SWIR OPDs,[17-18, 24-25, 34, 49-50] this BHJ shows a strong extinction coefficient $k$ up to $\lambda_g \approx 1050$ nm (**Figure 1a**, top figure; $\lambda_g$ is arbitrarily taken where $k = 0.25$). At $\lambda > \lambda_g$, $k$ reduces rapidly as $\lambda$ increases, and as a result the EQE value of reported PTB7-Th:COTIC-4F OPDs at $\lambda$ = 1100 nm are typically 10-15 %.[17-18, 24-25, 34, 49] In our RCE OPD, the thickness of BHJ film ($L_{BHJ}$) is chosen to target the resonance at the long-wavelength absorption tail ($\lambda > \lambda_g$, **Figure 1a,** bottom figure), where the absorption coefficient $k$ (blue curve, **Figure 1a**) is rather weak. At the resonance, an electromagnetic wave with a wavelength $\lambda = \lambda_m$ experiences constructive interference and the $\lambda_m$ photons are trapped in the device until they are absorbed by the active layer.[43, 47] Moreover, the broadband photoresponse window of OPD is not harmed by resonant-cavity mode, since the reflectance of one mirror is substantially low (**Figure S3**).[48, 51]

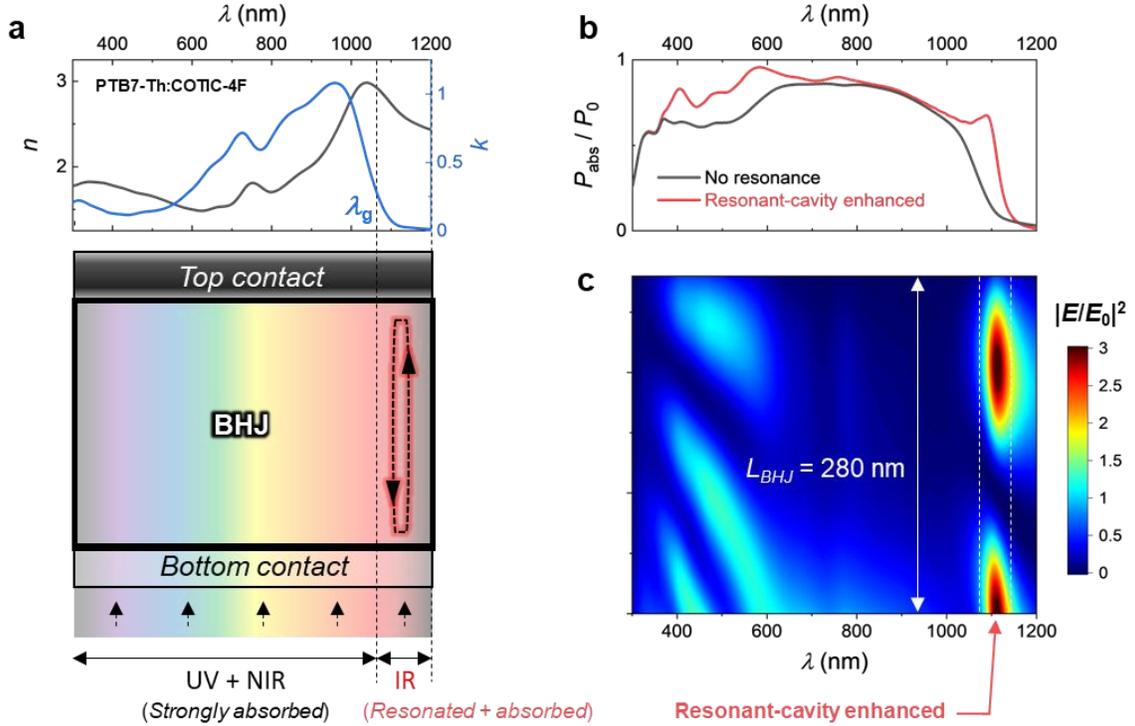

**Figure 1.** (a) *Top*: optical refractive index and extinction coefficient of PTB7-Th:COTIC-4F BHJ used for FDTD calculation. *Bottom*: Schematic of a broadband RCE OPD. This detector has a top reflective contact (MoO$_x$/Ag), and a transparent bottom contact (glass/ITO/ZnO). Photons with $\lambda < \lambda_g$ are strongly absorbed due to the high extinction coefficient of BHJ. Photons with $\lambda > \lambda_g$ are trapped

in the device due to resonance mode until they are absorbed. (b) *Black curve*: FDTD-calculated normalized power absorption ($P_{abs}/P_0$) of stand-alone 280 nm PTB7-Th:COTIC-4F BHJ. *Red curve:* with a sandwich of top and bottom contact, the same BHJ can absorb significantly more optical power at the long-wavelength absorption tail ($\lambda > \lambda_g$). (c) Time-average intensity map of local electric field $|E/E_0|^2$ at the cross-section of BHJ in the OPDs with thickness optimized for the resonance at $\lambda = 1100$ nm ($L_{BHJ} = 280$ nm).

The optical properties of a stand-alone BHJ ($L_{BHJ} = 280$ nm) and an OPD stack of glass/ITO (130 nm)/ZnO (40 nm)/BHJ (280 nm)/MoO$_x$ (7 nm)/Ag (100 nm) are simulated using FDTD method (see **Experimental Section** in the SI), and results are presented in **Figures 1b** and **S4**. The stand-alone BHJ film strongly absorbs photons when $\lambda < \lambda_g$, and when $\lambda > \lambda_g$, the normalized power absorption ($P_{abs}/P_0$) therefore drops rapidly as $\lambda$ increases (black curve, **Figures 1b**). However, an identical BHJ film sandwiched between the electrodes can support resonant cavity mode at $\lambda_m = 1100$ nm, which enhances $P_{abs}/P_0$ ratio at this wavelength by a factor of 5 (red curve, **Figures 1b**). The time-average intensity maps of the local electric field at the cross-section of a stand-alone BHJ film (**Figure S4b**) and a BHJ film in an OPD stack (**Figures 1c**) show the contrast of $|E/E_0|^2$ fields when there is no resonance and when resonant cavity mode is supported, respectively. Two strong local electric field centers at $\lambda_m = 1100$ nm can be observed in the map in **Figure 1c**, which corresponds to a resonance order of $m = 2$. By varying the thickness of the BHJ film, different resonance orders can be supported (**Figure S5**).

We survey the active layer thickness-dependent optical response of RCE OPDs. FDTD simulations are performed on glass/ITO (130 nm)/ZnO (40 nm)/BHJ/MoO$_x$ (7 nm)/Ag (100 nm) structure, with BHJ thicknesses ($L_{BHJ}$) varied from 20 nm to 900 nm. The calculated optical transmission and ideal EQE spectra are presented in **Figures 2a** and **2b**, respectively. In the transmission spectra map (**Figure 2a**), one can notice several "high-transmission strokes" at $\lambda \geq$ 1050 nm, which correspond to the resonance-mode order $m = 1$ to 4, as indicated by white arrows. At the same $\lambda$-$L_{BHJ}$ location but in the EQE spectra map (**Figure 2b**), enhanced-EQE peaks induced by resonance modes can be also located.

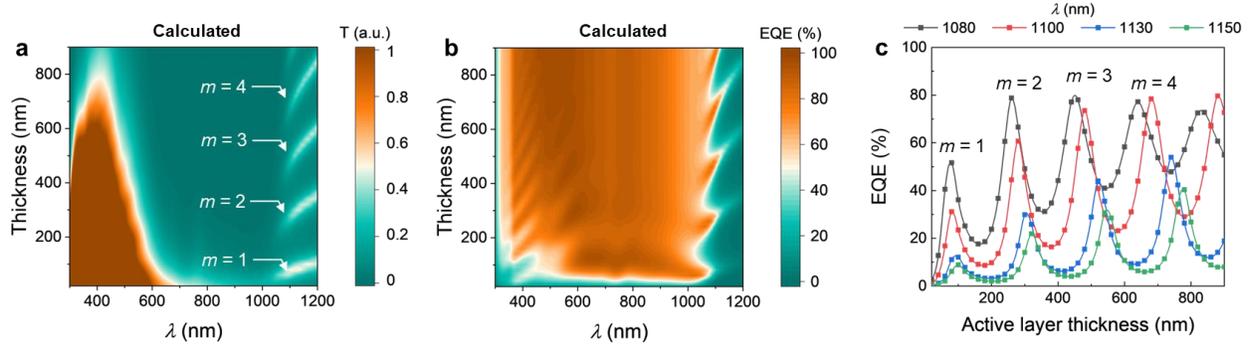

**Figure 2.** (a) FDTD-calculated optical transmission and (b) ideal EQE spectra of OPDs with active layer thickness varied from 20 nm to 900 nm. Resonance mode order from $m = 1$ to 4 for $\lambda \geq 1050$ nm are indicated in (a), and enhancements of EQE are also found at the same location in map (b). (c) EQE versus active layer thickness at different $\lambda$, extract from (b). The EQE peaks correspond to the enhanced-cavity resonance mode $m = 1$ to 4.

We extract the EQE versus the $L_{BHJ}$ at different wavelengths of $\lambda = $ 1080, 1100, 1130, and 1150 nm, and results are plotted in **Figure 2c**. Several evenly-spaced peaks corresponding to $m = $ 1 to 4 can be located and marked. The EQE value is greatly modulated when $L_{BHJ}$ increases, and the ratio between the maxima and minima EQE peak magnitude can be up to 5- to 10-fold. The strong dependency of EQEs on the film thicknesses can be attributed to the thickness-sensitive nature of the resonance modes.[30, 43] In contrast, at $\lambda \leq 1000$ nm where the extinction coefficient of the BHJ is fairly high and no resonance mode is induced, the EQE value is thickness-independent when $L_{BHJ} > 300$ nm (**Figure S6**). It is also worth noting that the resonance position $\lambda_m$ is also sensitive to the thicknesses of other layers in the stack such as ZnO and MoO$_x$, since the effective refractive index ($n_{eff}$) of the medium between two mirrors depends on these layers' thicknesses (**Figure S7b** and **S7d**). In contrast, varying the thickness of the top Ag electrode from 50 to 250 nm does not affect the resonance position, as the reflection of this layer only marginally changes when the thickness is over 50 nm (**Figure S7c**).

## 2.2. Experimental results

To experimentally demonstrate the concept, we use the PTB7-Th:COTIC-4F blend and fabricate the OPDs with the same architecture as in the simulation (**Figures 3a** and **3b**). Thickness variation of the active layer is done by changing the spin-speed and active layer solution concentration. More details on device fabrication can be found in the **Experimental Section**.

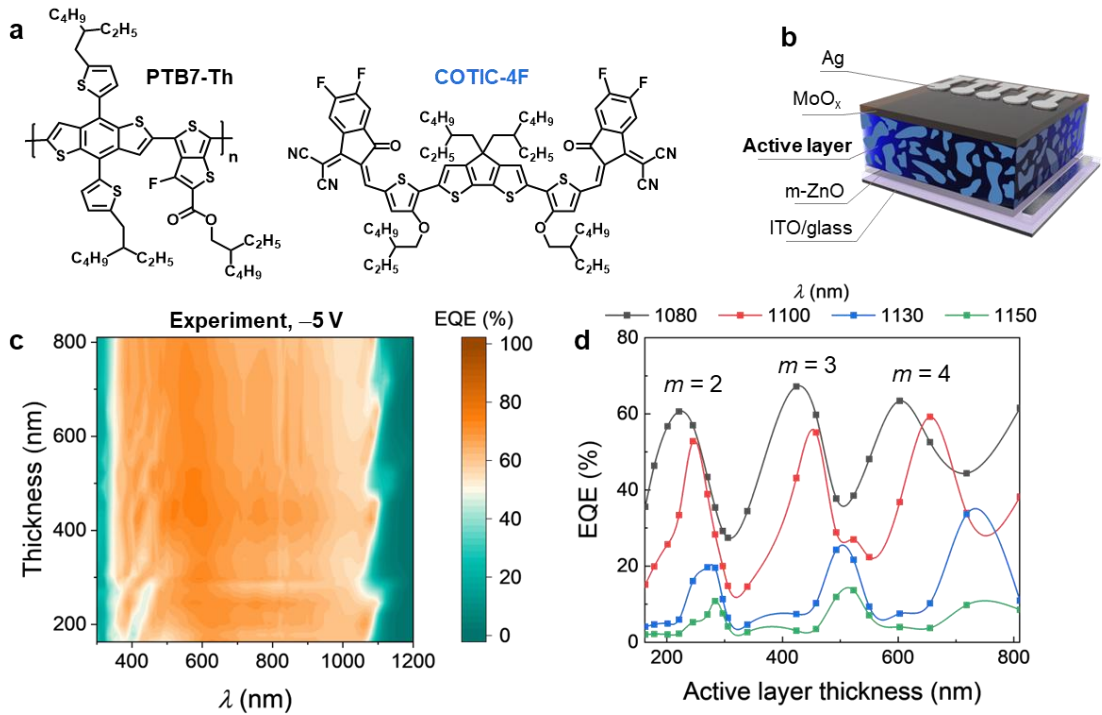

**Figure 3.** (a) Materials and (b) device architecture used for the experiment. (c) Experimental EQE spectra (−5 V) of OPDs with active layer thickness from 160 nm to 810 nm. (d) EQE (−5 V) versus

active layer thickness at different $\lambda$, extract from map (c).

The dark *J-V* characteristics of OPDs with increasing thicknesses (from 160 nm to 810 nm) are shown in **Figure S8a**. Unsurprisingly, we found that $J_d$ reduces when $L_{BHJ}$ increases. A similar observation is widely seen in the literature.[20] The thickness-dependent $J_d$ at −5 V are extracted and graphed (**Figure S8b**), manifesting that there is a threshold at $L_{BHJ}$ ~300-350 nm, where further increasing the thickness does not significantly reduce the dark current. **Figure 3c** displays the experimental EQE spectra (−5 V) of OPDs with the active layer thickness varied from 160 nm to 810 nm. We can locate the resonance modes $m$ = 2, 3, and 4 for $\lambda_m$ = 1100 nm at $L_{BHJ}$ = 245, 458, and 655 nm, respectively. The active layer thicknesses differences $\Delta L_{BHJ}$ between each resonance mode orders for $\lambda_m$ = 1100 nm is ~200-210 nm, which is fairly close to the FDTD-calculated value (200 nm). However, the BHJ thicknesses that support the resonance mode at $\lambda_m$ = 1100 nm (experimental values $L_{BHJ}$ = 245, 458, and 655 nm) are ~20-30 nm thinner than the simulated values ($L_{BHJ}$ = 280, 480, and 680 nm), which can be ascribed to the mismatches in the optical constant values used for simulation and imperfections in experimental devices (such as surface roughness, thickness variation of any layer in the stack, etc.).

We take a closer look at the EQEs at the resonance wavelengths as a function of $L_{BHJ}$, **Figure 3d**. Their thickness-dependent behaviors are similar to these of simulated results. Remarkably, the experimental EQE values (−5 V) at $\lambda_m$ = 1100 nm are 53 % ($m$ = 2) and 59 % ($m$ = 4), which are close to the ideal calculated EQE values (61 %, $m$ = 2 and 78 %, $m$ = 4), and this is a sign of good photon-to-electron conversion for this blend system even when the thickness is large. A high extraction coefficient at the resonance wavelength can be seen via bias-dependent EQE (**Figure S9a**). Charge collection ratio ($R_c$), estimated by $R_c$ = $EQE_{0V}/EQE_{-5V}$, are 0.87 ($m$ = 2) and 0.84 ($m = 4$) at $\lambda_m = 1100$ nm, which are already quite high without the assistance of the external bias. Sizable EQE enables very high shot-noise-limited specific detectivity $D^*_{sh}$ at low external applied voltage, as shot-noise reduces significantly to sub-nA/cm$^2$ level near 0 V reverse bias (**Figure S9a**). Interestingly, at a shorter wavelength range of $\lambda = 600 - 1000$ nm, $R_c$ reduces rapidly when $L_{BHJ}$ reaches ~500 nm or thicker (**Figures S9b** and **S9c**) due to space-charge collection.[18]

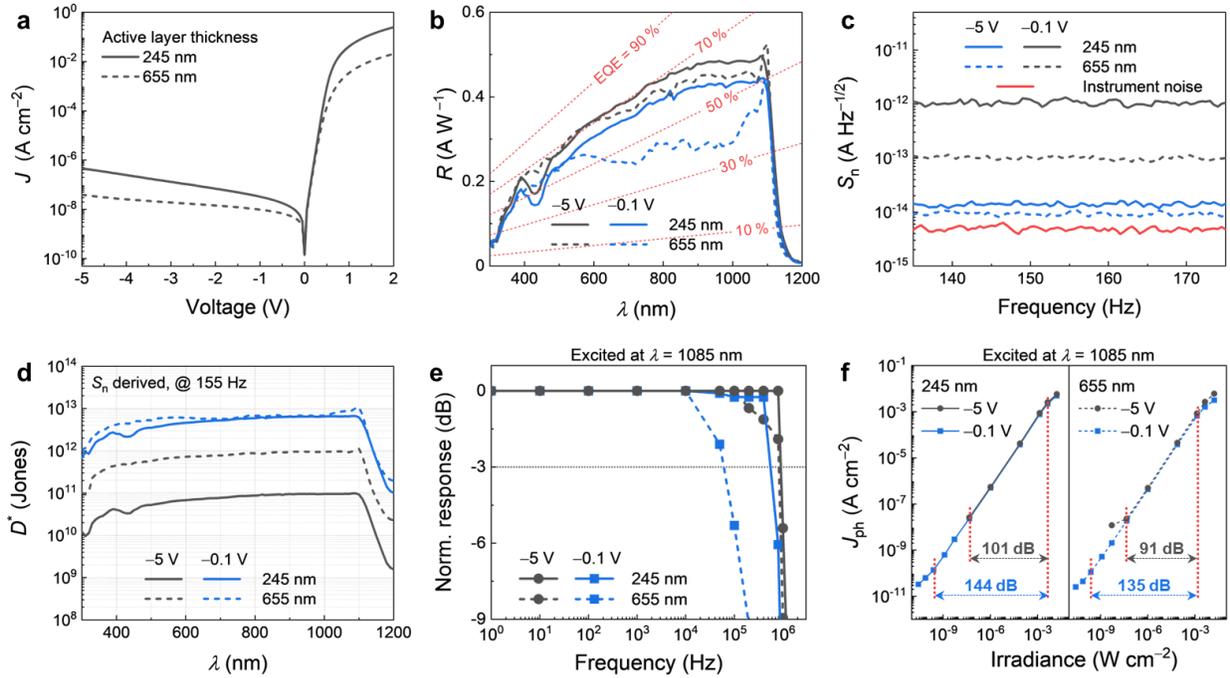

**Figure 4.** (a) Dark $J$-$V$ characteristic, (b) responsivity ($R$), (c) noise current spectral density ($S_n$) with a bandwidth of measurement $\Delta f = 1$ Hz, (d) specific detectivity ($D^*$) derived from $S_n$ at 155 Hz, (e) normalized response as a function of input signal frequency, and (f) linear dynamic range (LDR) of OPDs with an active layer thickness of 245 and 655 nm.

Two representative samples with $L_{BHJ} = 245$ and 655 nm (which support resonance modes $m = 2$ and 4 at $\lambda_m = 1100$ nm, respectively) are chosen for further standard photodetection characterizations. Atomic force microscopy (AFM) images of the two samples (**Figure S10**) suggest similar surface morphologies with the root-mean-square (rms) roughness ranging from 2.2 to 2.4 nm. These rms values are less than 1 % of the thicknesses of the BHJ films, which benefits reducing shunt leakage induced by nonuniformity of the interface. Grazing-incidence wide-angle X-ray scattering (GIWAXS) measurements suggest a preferred face-on orientation with no significant differences in terms of molecular ordering in both thin and thick film (**Figure S11**), which further supports that wavelength extension is attributed to variations of $L_{BHJ}$, rather than morphological properties.

**Figure 4a** shows a semilog plot of dark current-voltage characteristics of the OPDs. As $L_{BHJ}$ increases from 245 nm to 655 nm, the reverse $J_d$ at −5 V improves by an order of magnitude (456 nA/cm$^2$ to 38.2 nA/cm$^2$), while at $J_d$ at −0.1 V slightly reduces (5.5 nA/cm$^2$ to 3.0 nA/cm$^2$). The responsivity (*R*) spectra of the OPDs at different reverse biases are shown in **Figure 4b**. As aforementioned, both two OPDs are optimized for $\lambda_m = 1100$ nm, and at this wavelength, they achieved record high $R = 0.47$ and 0.52 A W$^{-1}$ at −5 V. To the best of our knowledge, this *R* value at $\lambda = 1100$ nm is about >3 times higher than a typical commercialized Si photodiode,[52] and >2 times higher than any binary OPDs in the literature.[11, 17, 21, 32, 34-35] Even at a small bias of −0.1 V, the photoresponse remains high as $R = 0.41$ and 0.45 A W$^{-1}$ for $L_{BHJ} = 245$ and 655 nm, at $\lambda = 1100$ nm, respectively. In terms of spectral shape, the *R* and EQE spectra of OPD with $L_{BHJ} = 245$ nm are overall "flat" under $\lambda = 400 - 1100$ nm range, with the shapes remaining the same regardless of reverse biases (**Figures S12a** and **S12b**). At a much thicker $L_{BHJ} = 655$ nm and no/low reverse bias of 0.1 V, we found a "valley" in the EQE spectra ranging from $\lambda = 600$ to 1000 nm (**Figures S12d**). This behavior might come from the space-charge collection narrowing as the charges generated by the middle wavelength range of 600-100 nm concentrate near the middle of the thick BHJ, and they have a higher chance to be recombined before extraction. However, charges can be extracted effectively with a sufficient small external bias of −1 V, as the EQE and *R* spectra shape of thick and thin devices are relatively similar at this reverse bias (**Figures S12b** and **S12d**). In addition, the EQE at $\lambda$ = 1100 nm of the thicker sample appears as a sharp peak, which implies a narrow window of resonant-enhance EQE, while in the thinner sample, only a small bump can be observed (see discussion of **Figures S2-3** for a detail explanation).

The noise assessment and specific detectivity are performed carefully to avoid performance overestimation. The noise current spectral density ($S_n$), consists of contributions from shot-noise ($S_{sh}$), thermal-noise ($S_{th}$), and flick-noise ($S_f$), is depicted in the below equation (with a bandwidth of measurement $\Delta f$ = 1 Hz),[17, 53]

$$S_n = \sqrt{S_{sh}^2 + S_{th}^2 + S_f^2} = \sqrt{2qI_d + \frac{4kT}{R_{sh}} + S_f^2}, \qquad (2)$$

where $R_{sh}$ represents the shunt resistance, $kT$ is the product of the Boltzmann constant ($k$) and the temperature ($T$), and $q$ is the elemental charge. For a photodetector with the device area of $A$, its specific detectivity ($D^*$) is defined as,

$$D^* = \frac{R\sqrt{A}}{S_n} \qquad (3)$$

The noise current spectral densities ($S_n$) of our OPDs are experimentally obtained using a fast Fourier transform (FFT) of the dark current at −0.1 V and −5 V, respectively (**Figure 4c**, more details can be found in SI, Experiment Section and **Figure S13**). The contribution of frequency-independent white-noises (*i.e.* shot-noise and thermal-noise) is also specified in **Table S1**. The instrument noise floor is consistently below $6\times10^{-15}$ A Hz$^{-1/2}$ (**Figure S13**), allowing accurate evaluation of the noise level of the OPDs. For $L_{BHJ}$ = 245 nm sample, the noise current densities at 155 Hz are around $1.4\times10^{-14}$ and $1.1\times10^{-12}$ A Hz$^{-1/2}$ at −0.1 V and −5 V, respectively. The values decrease to $9.2\times10^{-15}$ and $1.1\times10^{-13}$ A Hz$^{-1/2}$ at −0.1 V and −5 V, respectively, when the thickness increases to 655 nm. Using the measured $S_n$ at 155 Hz (which is also the frequency where the EQE spectra are collected, see **Experimental Section**), we obtained the specific detectivity $D^*$ using Equation (3), as illustrated in **Figure 4d**. At −0.1 V, OPDs with $L_{BHJ}$ = 245 and 655 nm exhibit peak $D^*$ of $1.1 \times 10^{13}$ and $6.2 \times 10^{12}$ Jones at $\lambda$ = 1100 nm, respectively, and stay > $10^{12}$ Jones over $\lambda$ = 300 – 1140 nm ranges. At −5 V, both OPDs suffer from high flicker-noises and high $S_n$ is observed (**Table S1**), however, $D^*$ remains ~$10^{12}$ Jones at $\lambda$ = 1100 nm for a thick device. These obtained values of $D^*_{sh}$ and $D^*$ place PTB7-Th:COTIC-4F RCE OPDs among the best OPDs for $\lambda \geq 1080$ nm in the literature. We will discuss this comparison in more detail in a later part of the manuscript.

Utilizing resonant-optical cavity in inorganic photodetectors[48, 51] and CT-state-based OPDs[30] is known for not impairing response time (rise/fall time) as well as $f_{3dB}$ (the frequency at which the output of a detector is attenuated to ~70 % of original amplitude) of the device. Particularly in this case, FDTD simulations reveal that the optical resonance is already well-defined in the ps time scale after light excitation. This time scale is much faster than the response times of a typical OPD (μs time scale) and therefore no negative effect from the RCE structure on the response speed is expected. To evaluate the speed of the photodetector, the transient photoresponse behaviors under the illumination of an IR light source (center wavelength $\lambda$ = 1085 nm) are collected. As shown in **Figure S14a**, the OPD with $L_{BHJ}$ = 245 nm exhibits sub-μs response times even at a small bias of −0.1 V. The OPD with $L_{BHJ}$ = 655 nm requires a much larger bias of −5 V to get sub-µs response times, and at −0.1 V, both the rise and fall time are above 10 µs (**Figure S14b**). The rapid on-off switching times allow the OPDs to follow an optical square-wave at a high modulating frequency (**Figure S15**). By extracting the normalized photoresponse as a function of the input signal frequency, we obtain the cutoff frequency ($f_{3dB}$) of ~900 kHz (−5 V) and ~550 kHz (−0.1 V) for $L_{BHJ}$ = 245 nm sample and ~820 kHz (−5 V) and ~60 kHz (−0.1 V) for $L_{BHJ}$ = 655 nm sample (**Figure 4e** and **Figure S16**), which show a consistent trend seen with the response time data. Rapid response time and wide bandwidth of this sensing platform are suitable for high-speed image sensing applications[54] and optical communications.[55-56]

Besides the sensor speed and applicable bandwidth, the linearity of a photodetector, measured by linear dynamic range (LDR), is a key parameter for applications such as image sensors. LDR refers to the range of light intensity levels over which photocurrent and light intensity shows a linear relation in a log–log plot. Typically, the LDR can be calculated by,

$$\text{LDR} = 20 \times \log \frac{J_{\text{upper}}}{J_{\text{lower}}} \text{ (dB)}, \qquad (4)$$

where $J_{\text{upper}}$ and $J_{\text{lower}}$ are the photocurrent below and beyond which the photocurrent density ($J_{\text{ph}}$) – irradiance becomes non-linear. As shown in **Figure 4f**, $J_{\text{ph}}$ under irradiation of 1085 nm light of different intensities are presented. At −0.1 V, we obtain LDR of 144 dB and 133 dB for the thin and the thick devices, respectively. At −5 V, the LDRs of both OPDs reduce to about 100 dB, due to the large dark current background which limits the $J_{\text{lower}}$ to >10 nA/cm$^2$ level.

Summarily, by simply matching the active layer thickness to optical resonant conditions of the "cavity" formed by the top and bottom contacts, a PTB7-Th:COTIC-4F OPD under −0.1 V can obtain an enhanced-EQE of 50 % and $D^*_{sh} > 10^{13}$ Jones at $\lambda = 1100$ nm, while maintains sub-µs response time, $f_{3dB} > 500$ kHz, and LDR > 140 dB. Importantly, this approach can be straightforwardly applied to any BHJ system and inverted/conventional device architecture. In the following section, we showcase an example of using this strategy in reaching a sizable EQE at an even further wavelength range of $\lambda = 1150\text{-}1200$ nm.

## 2.3. Universality of the approach

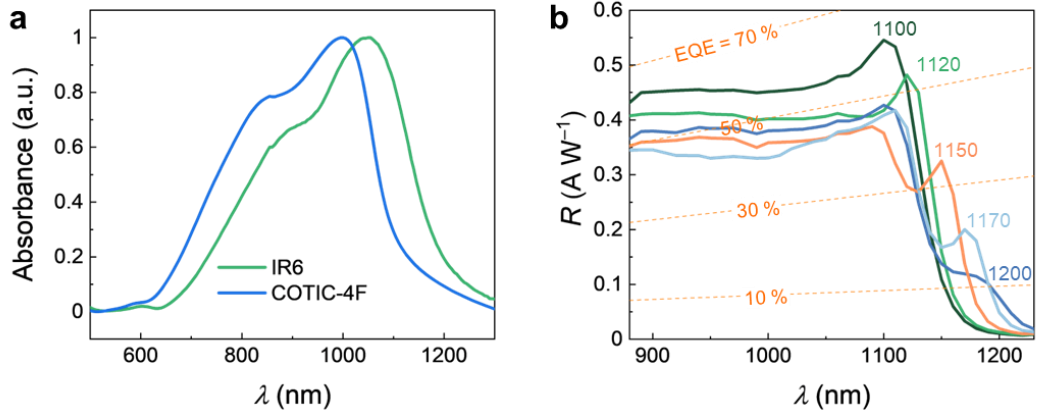

**Figure 5.** (a) Thin film absorbance spectra of IR6 and COTIC-4F NFAs. (b) Responsivity ($R$) spectra of PTB7-Th:IR6 OPDs with different $L_{BHJ}$ optimized for $\lambda = 1100$, 1120, 1150, 1170, and 1200 nm.

In order to furtherly extend the spectral sensitivity into the IR range, we employ a novel NFA IR6 with a red-shifted thin film absorbance peak compared to COTIC-4F ($\Delta\lambda \approx 50$ nm, **Figure 5a**). IR6 has acceptor units with stronger electron-withdrawing nature than COTIC-4F has. These acceptor units decrease optical bandgap and increase ionization potential at the same time, enabling the red-shift with sufficient hole transfer rate from excited NFA to PTB7-Th.

Remarkably, the EQE spectra of PTB7-Th:IR6 OPD covers a wide spectrum from 300 nm up to 1180 nm, with comparable EQE magnitudes to these of PTB7-Th:COTIC-4F OPD (**Figure**

**S18**). As expected, thickness-dependent EQEs at sub-bandgap wavelengths exhibit several resonance modes, with consecutive modes spacing roughly 200-250 nm (**Figure S19**). Representative $R$ spectra (−5 V) of OPDs with $L_{BHJ}$ optimized for $\lambda = 1100 - 1200$ nm are presented in **Figure 5b**. At $\lambda = 1100, 1120, 1150, 1170,$ and 1200 nm, $R = 0.55, 0.48, 0.32, 0.20,$ and 0.08 A W$^{-1}$ at −5 V, respectively. Even at 0 V, photoresponses at respective $\lambda$ remain high as $R = 0.43, 0.37, 0.19, 0.09,$ and 0.05 A W$^{-1}$, despite thick BHJs being used. It is worth noting that the dark current densities ($J_d$) of PTB7-Th:IR6 OPDs are only marginally higher than PTB7-Th:COTIC-4F OPDs at a similar thickness (**Figure S20**), and are a few orders of magnitude lower than these of OPDs based on ultra-narrow bandgap polymer.[37, 40, 57] Therefore, the specific detectivity of PTB7-Th:IR6 OPDs remains high (**Figures S21-22**). In addition, fast detection response (< 10 μs at −5 V), high cut-off frequencies ($f_{3dB} > 100$ kHz at −5 V) and wide LDR (> 130 dB at −5 V) can be achieved with PTB7-Th:IR6 OPDs. These characterizations can be found in the SI, **Figures S21-S26**.

As the resonant-enhanced EQEs are universally applicable in different BHJ systems with regular/inverted photodiode architectures, it is worth echoing the potential artifacts that might arise in analysis using the long-wavelength tail of EQE spectra. Organic solar cells/photodetectors are stacks of thin films, which can form low finesse cavities and produce "bumps" on the sub-bandgap photovoltaic EQE spectra. These bumps have been experimentally shown to emerge in the EQE range of $10^2 - 10^0$ (%) in this work, and down to $10^{-1} - 10^{-7}$ (%) if a highly-sensitive EQE measurement were employed.[58-61] Optical interference effects induced by resonant-cavity can produce akin features which can be confused with CT-state,[58-59] trap states,[60-62] and can interfere with Urbach energy[15] determination.

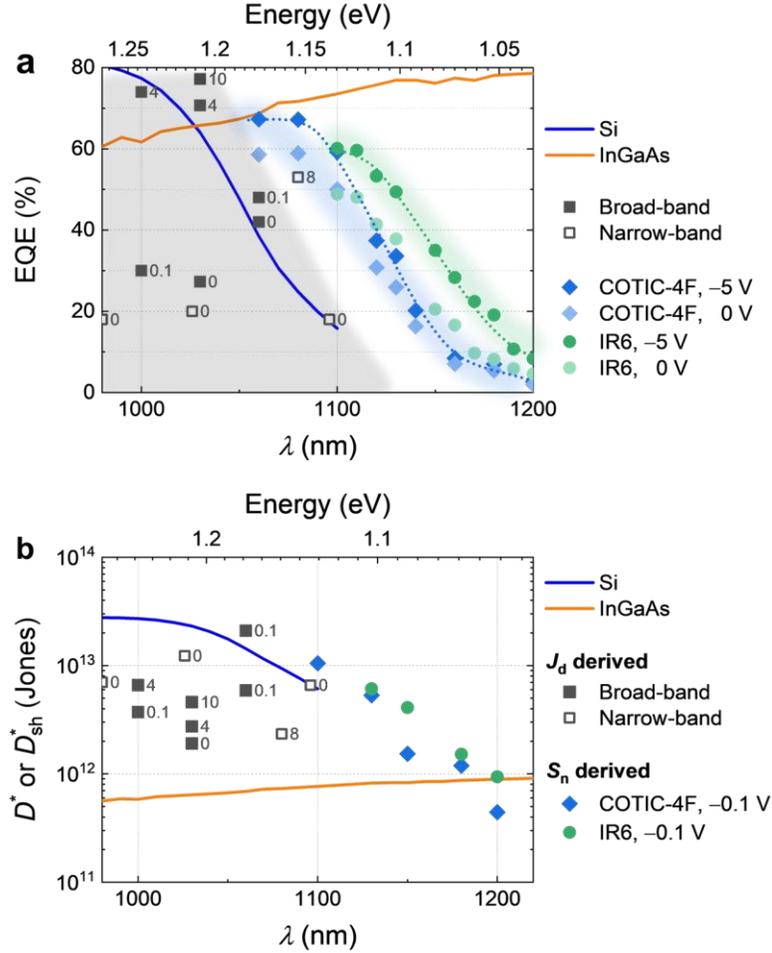

**Figure 6.** (a) EQE and (b) specific detectivity (derived from dark current density $J_d$ or measured noise current $S_n$) of OPDs using NFA as a SWIR absorber with EQE ≥ 20 %, covering the wavelength range of $\lambda = 980 - 1200$ nm. The small numbers to the right of each literature data point (solid/open black square symbols) are the reversed biases at which the OPDs were reported. Our RCE OPDs based on PTB7-Th:COTIC-4F and PTB7-Th:IR6 are among the first NFA-based OPDs reaching EQE > 30 % at $\lambda = 1100 - 1150$ nm.

**Figure 6** visualizes the state-of-the-art performance of NFA-based OPDs with $\lambda = 980 - 1200$ nm in the literature, in comparison with commercialized inorganic photodetectors (such as Si and InGaAs) and our RCE-based OPDs (data table and references in **Figure 6** can be found in **Table S2**). Using PTB7-Th:IR6, we successfully extend the spectral sensitivity to a longer wavelength of ~100 nm in comparison to a commercial Si PD (**Figure 6a**). In particular, the EQEs of RCE OPDs are 67, 60, 49, 35, 19, and 8 % (−5 V) at $\lambda$ = 1080, 1100, 1130, 1150, 1180, and 1200 nm, respectively. Even without an applied bias, EQEs of 59, 50, 38, 21, 8, and 5 % at $\lambda$ = 1080, 1100, 1130, 1150, 1180, and 1200 nm are obtained, respectively, which surpass the EQE of the existing organic photodiodes in the range of $\lambda$ = 1080 – 1150 nm, to the best of our knowledge. Low dark current density $J_d$ and noise current $S_n$ enable high specific detectivity in the range of $\lambda \leq$ 1200 nm (**Figure 6b**), which can be comparable to InGaAs photodetector.

## 3. Conclusion

In summary, we demonstrate a simple approach to expand the detection range of OPDs to longer wavelengths, by exploiting the low finesse cavity effect in the photodiode stack. By fine-tuning the thickness of the PTB7-Th:COTIC-4F active layer, a record sensitivity ($R$ = 0.52 A W$^{-1}$) and specific detectivity ($D^*$ = 1.1 × 10$^{13}$ Jones) can be obtained at $\lambda$ = 1100 nm, while broadband wavelength spectra, bandwidth, speed, and linearity are preserved. By employing a novel NFA IR6 with a red-shifted absorbance peak compared to COTIC-4F, we successfully achieve an EQE ~ 40 % at $\lambda$ = 1150 nm and ~20 % at $\lambda$ = 1180 nm while maintaining $J_d$ well below 100 nA cm$^{-2}$ at −5 V. This study highlights the significance of optical design in optoelectronic devices, offering a notably simpler approach to extend the photodetection range compared to conventional methods that rely on developing absorbers with extremely low optical gaps.

**Acknowledgments**

This project is supported by the Mitsubishi Chemical Center for Advanced Materials (MC-CAM). The NFA IR6 in this work is provided by Mitsubishi Chemical Corporation (MCC). The authors gratefully thank Dr. Shen Xing and Prof. Karl Leo (Technische Universität Dresden, Germany) for the discussion to improve the noise current measurement setup.

**Author contributions**

H.L. and T.-Q. N. conceived the idea. H.L. designed the experiments, performed FDTD calculations, carried out device fabrication and characterizations, and wrote the manuscript. C.K. and A.Y. supported device optimization and characterizations. A.Y., S.C., and B.M.K. conducted AFM, GIWAXS, UV-VIS, thickness measurements, and data analysis. P.P. synthesized COTIC-4F NFA. Y.M. and H.N. synthesized the IR6 NFA. T.-Q.N. was responsible for project planning, data interpretation, group project management, and manuscript writing. All authors gave feedback on the final paper.

**References**

[1] F. Cao, L. Liu, L. Li, *Materials Today* **2022**, 62, 327.
[2] X. Guan, X. Yu, D. Periyanagounder, M. R. Benzigar, J. K. Huang, C. H. Lin, J. Kim, S. Singh, L. Hu, G. Liu, D. Li, J. H. He, F. Yan, Q. J. Wang, T. Wu, *Adv. Opt. Mater.* **2020**, 9, 2001708.
[3] E. Hemmer, N. Venkatachalam, H. Hyodo, A. Hattori, Y. Ebina, H. Kishimoto, K. Soga, *Nanoscale* **2013**, 5, 11339.
[4] E. Hemmer, A. Benayas, F. Légaré, F. Vetrone, *Nanoscale Horizons* **2016**, 1, 168.
[5] N. Li, X. Hu, X. Sui, Q. Chen, T. N. Ng, *ACS Applied Electronic Materials* **2023**.
[6] Z. Wu, Y. Zhai, H. Kim, J. D. Azoulay, T. N. Ng, *Acc. Chem. Res.* **2018**, 51, 3144.
[7] J. Liu, P. Liu, D. Chen, T. Shi, X. Qu, L. Chen, T. Wu, J. Ke, K. Xiong, M. Li, *Nature Electronics* **2022**, 5, 443.
[8] P. Wang, H. Xia, Q. Li, F. Wang, L. Zhang, T. Li, P. Martyniuk, A. Rogalski, W. Hu, *Small* **2019**, 15, 1904396.
[9] A. Rogalski, P. Martyniuk, M. Kopytko, W. Hu, *Applied Sciences* **2021**, 11, 501.
[10] S. Xing, V. C. Nikolis, J. Kublitski, E. Guo, X. Jia, Y. Wang, D. Spoltore, K. Vandewal, H. Kleemann, J. Benduhn, *Adv. Mater.* **2021**, 33, 2102967.
[11] J. H. Kim, A. Liess, M. Stolte, A. M. Krause, V. Stepanenko, C. Zhong, D. Bialas, F. Spano, F. Würthner, *Adv. Mater.* **2021**, 33, 2100582.
[12] W. Yang, W. Qiu, E. Georgitzikis, E. Simoen, J. Serron, J. Lee, I. Lieberman, D. Cheyns, P. Malinowski, J. Genoe, *ACS Appl. Mater. Interfaces* **2021**, 13, 16766.
[13] H. Jinno, T. Yokota, M. Koizumi, W. Yukita, M. Saito, I. Osaka, K. Fukuda, T. Someya,

*Nat. Commun.* **2021**, 12, 1.

[14] T. Yokota, P. Zalar, M. Kaltenbrunner, H. Jinno, N. Matsuhisa, H. Kitanosako, Y. Tachibana, W. Yukita, M. Koizumi, T. Someya, *Science advances* **2016**, 2, e1501856.
[15] C. Kaiser, O. J. Sandberg, N. Zarrabi, W. Li, P. Meredith, A. Armin, *Nat. Commun.* **2021**, 12, 1.
[16] B. Park, J. Jung, D. H. Lim, H. Lee, S. Park, M. Kyeong, S. J. Ko, S. H. Eom, S. H. Lee, C. Lee, *Advanced Functional Materials* **2022**, 32, 2108026.
[17] Y. Song, Z. Zhong, P. He, G. Yu, Q. Xue, L. Lan, F. Huang, *Adv. Mater.* **2022**, 2201827.
[18] Q. Liu, S. Zeiske, X. Jiang, D. Desta, S. Mertens, S. Gielen, R. Shanivarasanthe, H.-G. Boyen, A. Armin, K. Vandewal, *Nat. Commun.* **2022**, 13, 1.
[19] J. Huang, J. Lee, H. Nakayama, M. Schrock, D. X. Cao, K. Cho, G. C. Bazan, T.-Q. Nguyen, *ACS Nano* **2021**, 15, 1753.
[20] J. Huang, J. Lee, J. Vollbrecht, V. V. Brus, A. L. Dixon, D. X. Cao, Z. Zhu, Z. Du, H. Wang, K. Cho, *Adv. Mater.* **2020**, 32, 1906027.
[21] J. L. Wu, L. H. Lai, Y. T. Hsiao, K. W. Tsai, C. M. Yang, Z. W. Sun, J. C. Hsieh, Y. M. Chang, *Adv. Opt. Mater.* **2022**, 10, 2101723.
[22] L.-H. Lai, C.-C. Hsieh, J.-L. Wu, Y.-M. Chang, *ACS Applied Electronic Materials* **2021**, 4, 168.
[23] W.-L. Li, C.-H. Hou, C.-M. Yang, K.-W. Tsai, J.-L. Wu, Y.-T. Hsiao, C. Hanmandlu, C.-W. Chu, C.-H. Tsai, C.-Y. Liao, *Journal of Materials Chemistry A* **2021**, 9, 17967.
[24] J. Lee, S.-J. Ko, H. Lee, J. Huang, Z. Zhu, M. Seifrid, J. Vollbrecht, V. V. Brus, A. Karki, H. Wang, K. Cho, T.-Q. Nguyen, G. C. Bazan, *ACS Energy Letters* **2019**, 4, 1401.
[25] J. Lee, S. J. Ko, M. Seifrid, H. Lee, B. R. Luginbuhl, A. Karki, M. Ford, K. Rosenthal, K. Cho, T. Q. Nguyen, *Advanced Energy Materials* **2018**, 8, 1801212.
[26] J. Lee, S. J. Ko, M. Seifrid, H. Lee, C. McDowell, B. R. Luginbuhl, A. Karki, K. Cho, T. Q. Nguyen, G. C. Bazan, *Advanced Energy Materials* **2018**, 8, 1801209.
[27] M. Biele, C. Montenegro Benavides, J. Hürdler, S. F. Tedde, C. J. Brabec, O. Schmidt, *Advanced Materials Technologies* **2019**, 4, 1800158.
[28] X. Ma, H. Bin, B. T. van Gorkom, T. P. A. van der Pol, M. J. Dyson, C. H. L. Weijtens, M. Fattori, S. C. J. Meskers, A. van Breemen, D. Tordera, R. A. J. Janssen, G. H. Gelinck, *Adv. Mater.* **2022**, DOI: 10.1002/adma.202209598e2209598.
[29] N. Strobel, M. Seiberlich, T. Rödlmeier, U. Lemmer, G. Hernandez-Sosa, *ACS Appl. Mater. Interfaces* **2018**, 10, 42733.
[30] Z. Tang, Z. Ma, A. Sánchez‐Díaz, S. Ullbrich, Y. Liu, B. Siegmund, A. Mischok, K. Leo, M. Campoy‐Quiles, W. Li, *Adv. Mater.* **2017**, 29, 1702184.
[31] S. Ullbrich, B. Siegmund, A. Mischok, A. Hofacker, J. Benduhn, D. Spoltore, K. Vandewal, *J. Phys. Chem. Lett.* **2017**, 8, 5621.
[32] Z. Zhong, F. Peng, L. Ying, G. Yu, F. Huang, Y. Cao, *Science China Materials* **2021**, 64, 2430.
[33] S. Deng, L. Zhang, J. Zheng, J. Li, S. Lei, Z. Wu, D. Yang, D. Ma, J. Chen, *Adv. Opt. Mater.* **2022**, 2200371.
[34] C. Xu, P. Liu, C. Feng, Z. He, Y. Cao, *Journal of Materials Chemistry C* **2022**, 10, 5787.
[35] K.-W. Tsai, G. Madhaiyan, L.-H. Lai, Y.-T. Hsiao, J.-L. Wu, C.-Y. Liao, C.-H. Hou, J.-J. Shyue, Y.-M. Chang, *ACS Appl. Mater. Interfaces* **2022**, 14, 38004.
[36] N. Li, N. Eedugurala, J. D. Azoulay, T. N. Ng, *Cell Reports Physical Science* **2022**, 3, 100711.

[37] I. Park, C. Kim, R. Kim, N. Li, J. Lee, O. K. Kwon, B. Choi, T. N. Ng, D. S. Leem, *Adv. Opt. Mater.* **2022**, 10, 2200747.
[38] F. Verstraeten, S. Gielen, P. Verstappen, J. Raymakers, H. Penxten, L. Lutsen, K. Vandewal, W. Maes, *Journal of Materials Chemistry C* **2020**, 8, 10098.
[39] S. Gielen, C. Kaiser, F. Verstraeten, J. Kublitski, J. Benduhn, D. Spoltore, P. Verstappen, W. Maes, P. Meredith, A. Armin, *Adv. Mater.* **2020**, 32, 2003818.
[40] Z. Wu, W. Yao, A. E. London, J. D. Azoulay, T. N. Ng, *Advanced Functional Materials* **2018**, 28, 1800391.
[41] A. E. London, L. Huang, B. A. Zhang, M. B. Oviedo, J. Tropp, W. Yao, Z. Wu, B. M. Wong, T. N. Ng, J. D. Azoulay, *Polymer Chemistry* **2017**, 8, 2922.
[42] Z. Wu, N. Li, N. Eedugurala, J. D. Azoulay, D.-S. Leem, T. N. Ng, *npj Flexible Electronics* **2020**, 4, 1.
[43] B. Siegmund, A. Mischok, J. Benduhn, O. Zeika, S. Ullbrich, F. Nehm, M. Bohm, D. Spoltore, H. Frob, C. Korner, K. Leo, K. Vandewal, *Nat. Commun.* **2017**, 8, 15421.
[44] C. Kaiser, K. S. Schellhammer, J. Benduhn, B. Siegmund, M. Tropiano, J. Kublitski, D. Spoltore, M. Panhans, O. Zeika, F. Ortmann, P. Meredith, A. Armin, K. Vandewal, *Chemistry of Materials* **2019**, 31, 9325.
[45] Y. Wang, B. Siegmund, Z. Tang, Z. Ma, J. Kublitski, S. Xing, V. C. Nikolis, S. Ullbrich, Y. Li, J. Benduhn, D. Spoltore, K. Vandewal, K. Leo, *Adv. Opt. Mater.* **2020**, 9, 2001784.
[46] J. Vanderspikken, Q. Liu, Z. Liu, T. Vandermeeren, T. Cardeynaels, S. Gielen, B. Van Mele, N. Van den Brande, B. Champagne, K. Vandewal, *Advanced Functional Materials* **2022**, 32, 2108146.
[47] J. Vanderspikken, W. Maes, K. Vandewal, *Advanced Functional Materials* **2021**, 31, 2104060.
[48] K. Kishino, M. S. Unlu, J.-I. Chyi, J. Reed, L. Arsenault, H. Morkoc, *IEEE Journal of Quantum Electronics* **1991**, 27, 2025.
[49] J. Huang, J. Lee, M. Schrock, A. L. Dixon, A. T. Lill, K. Cho, G. C. Bazan, T.-Q. Nguyen, *Materials Horizons* **2020**, 7, 3234.
[50] H. M. Luong, S. Chae, A. Yi, K. Ding, J. Huang, B. M. Kim, C. Welton, J. Chen, H. Wakidi, Z. Du, H. J. Kim, H. Ade, G. N. M. Reddy, T.-Q. Nguyen, *ACS Energy Letters* **2023**, 8, 2130.
[51] M. S. Ünlü, S. Strite, *J. Appl. Phys.* **1995**, 78, 607.
[52] Thorlab photodetectors, https://www.thorlabs.com/newgrouppage9.cfm?objectgroup_id=285, accessed.
[53] Y. Fang, A. Armin, P. Meredith, J. Huang, *Nat. Photonics* **2019**, 13, 1.
[54] K. Nishimura, S. Shishido, Y. Miyake, H. Kanehara, Y. Sato, J. Hirase, Y. Sato, Y. Tomekawa, M. Yamasaki, M. Murakami, M. Harada, Y. Inoue, *Jpn. J. Appl. Phys.* **2018**, 57, 1002B4.
[55] T. Mueller, F. Xia, P. Avouris, *Nat. Photonics* **2010**, 4, 297.
[56] A. Morteza Najarian, M. Vafaie, A. Johnston, T. Zhu, M. Wei, M. I. Saidaminov, Y. Hou, S. Hoogland, F. P. García de Arquer, E. H. Sargent, *Nature Electronics* **2022**, 5, 511.
[57] J. Yang, J. Huang, R. Li, H. Li, B. Sun, Q. Lin, M. Wang, Z. Ma, K. Vandewal, Z. Tang, *Chemistry of Materials* **2021**, 33, 5147.
[58] A. Armin, M. Velusamy, P. Wolfer, Y. Zhang, P. L. Burn, P. Meredith, A. Pivrikas, *ACS Photonics* **2014**, 1, 173.
[59] G. F. Burkhard, E. T. Hoke, M. D. McGehee, *Adv. Mater.* **2010**, 22, 3293.

[60] C. Kaiser, S. Zeiske, P. Meredith, A. Armin, *Adv. Opt. Mater.* **2020**, 8, 1901542.
[61] A. Armin, N. Zarrabi, O. J. Sandberg, C. Kaiser, S. Zeiske, W. Li, P. Meredith, *Advanced Energy Materials* **2020**, 10, 2001828.
[62] C. Kaiser, O. Sandberg, S. Zeiske, S. Gielen, W. Maes, K. Vandewal, P. Meredith, A. Armin, **2021**.

# Supplementary Information

## Experimental Section

### Materials

The full names of the materials used are as follows:

PTB7-Th: poly[4,8-bis(5-(2-ethylhexyl)thiophen-2-yl)benzo[1,2-b;4,5-b']dithiophene-2,6-diyl-alt-(4-(2-ethylhexyl)-3-fluorothieno[3,4-b]thiophene-)-2-carboxylate-2-6-diyl)]

COTIC-4F: 2,2'-((2Z,2'Z)-((5,5'-(4,4-bis(2-ethylhexyl)-4H-cyclopenta[1,2-b:5,4-b']dithiophene-2,6- diyl)bis(4-((2-ethylhexyl)oxy)thiophene-5,2-diyl))bis(methanylylidene))bis(5,6-difluoro-3-oxo-2,3-dihydro-1H-indene-2,1-diylidene))dimalononitrile

All reagents were used as received without further purification. Zinc acetate dihydrate $(Zn(CH_3COO)_2 \cdot 2H_2O)$ was purchased from Honeywell. 2-methylimidazole was purchased from TCI chemical. 2-methoxyethanol and ethanolamine were purchased from Sigma-Aldrich. N,N-Dimethylformamide was purchased from EMD Millipore. The donor polymer, PTB7-Th, was purchased from 1-Materials (Lot No. SX-8015A, molecular weight ~120k). The non-fullerene acceptor (NFA) COTIC-4F was synthesized following a literature procedure.[1] $MoO_x$ and silver (Ag) pallets for thermal evaporation were obtained from Sigma-Alrich and Kurt J. Lesker company, respectively.

### Device fabrications

Indium tin oxide (ITO) substrates (~130 nm, ≤15 Ω/sq) were cleaned by scrubbing with a commercial detergent, then ultrasonicated in deionized water, acetone, and 2-propanol for 20 min, respectively. The cleaned substrates were dried in an oven at 130 °C for 30 minutes and treated under a UV-ozone lamp for 20 minutes. The sol-gel precursor solution for ZnO was prepared by mixing 100 mg of zinc acetate dihydrate and 60 μL of ethanolamine in 1 mL of 2-methoxyethanol, followed by stirring at 60 °C for 1 h. Immediately after the ozone treatment, sol-gel precursor solution for ZnO was spin-casted onto an ITO substrate at 2000 rpm for 30 s, followed by annealing at 200 °C for 1 h. After the ZnO film was formed and cooled down, we modified the surface of ZnO (m-ZnO) using 2-methylimidazole. Firstly, 5 mg of 2-methylimidazole was dissolved in dimethylformamide (DMF). Then, the solution was spin-coated onto the ZnO film with a spin speed of 2000 rpm for 60 s, followed by annealing at 80 °C for 1 h. The substrates of ITO/m-ZnO are loaded into a glovebox filled with nitrogen for the spin-casting of the active layer. The active layer solutions of PTB7-Th:NFA (1:1.5 weight ratio) in chlorobenzene:1-chloronaphthalene (98:2, v/v) at different concentrations were stirred overnight (35 mg/ml, stirred at room temperature and 45 mg/ml, stirred at 40 °C) and spun-coated at different speeds (500 to 1500 rpm for 2 minutes) to obtained varied thicknesses ranging from 160 nm to 810 nm. All the substrates were then soft-baked at 80 °C for 5 minutes to remove the residual solvent. Afterward, the substrates were loaded into a high vacuum chamber (< $10^{-6}$ torr), and 7 nm $MoO_x$ and 100 nm Ag were thermally evaporated to form the hole transporting layer and the top electrode. The effective device area is 4.7 $mm^2$.

**Finite-difference time domain (FDTD) method simulations**

FDTD calculations of OPD stacks were carried out using commercial software (Lumerical FDTD Solutions).[2] 2D simulation region was indicated as in **Figure S1** with optical parameters of ITO, ZnO, $MoO_x$, and Ag taken from ref. [3].; optical parameters of the active layer were experimentally obtained by optical ellipsometry as shown in **Figure 1a**. The refractive index of glass was set to 1.5. The mesh size of 5 nm × 1 nm was chosen. The idea EQE spectra were modeled based on the methodology shown in ref. [4]. We assume that each photon absorbed by the active layer will generate an electron-hole pair and the extraction rate was set at 100 % (no loss). A 2D frequency-domain profile and power monitor (yellow rectangle, **Figure S1**) recorded the photon absorption as well as the electric field profile at the cross-section of the active layer. The optical transmission of the OPD was captured by another 1D frequency-domain profile and power monitor (orange line, **Figure S1**) set near the top of the simulation box.

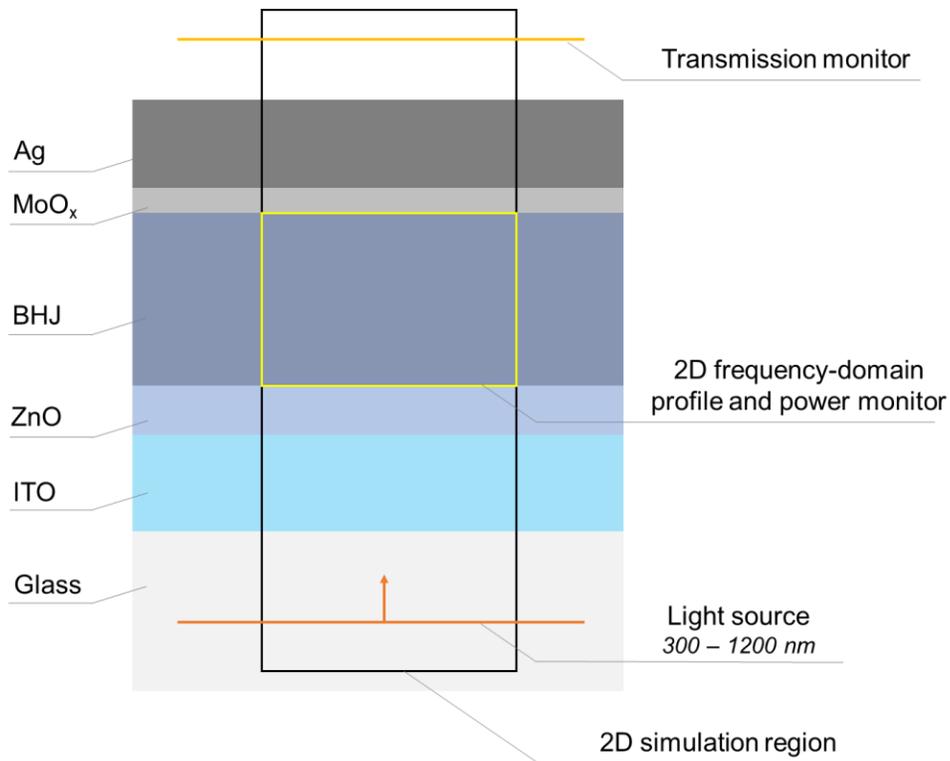

**Figure S1.** A schematic of device structure and finite-difference time domain (FDTD) simulation setup.

**Device characterizations**

The current-voltage (*J-V*) characteristics were measured by a Keithley 4200 semiconductor characterization system (SCS). The external quantum efficiency (EQE) spectra were taken with a custom-built setup including a 75-watt Xenon light source coupled with a monochromator and an optical chopper (chopping frequency was set at 155 Hz). A reference Si photodetector (NIST-calibrated Newport 818-UV Si photodiode) and a Ge photodetector (NIST-calibrated Newport 818-IR Ge photodiode) were used for calibration. The photocurrent was amplified using an SR570 low-noise pre-amplifier (Stanford Research System) and the output voltage signals were measured using an SR810 lock-in amplifier (Stanford Research System).

For linear dynamics range (LDR), $f_{3dB}$ cut-off frequency, and rise/fall time measurements, light-emitting diodes with an emission center wavelength of 1085 nm (LED1085L, Thorlabs) were used to illuminate the samples. For LDR measurement, the irradiance-dependent photocurrents from the OPDs were collected by Keithley 4200-SCS. For $f_{3dB}$ cut-off frequency and rise/fall time measurements, the light sources were modulated at different frequencies by a DS345 function generator (Stanford Research Systems). The output photocurrent from OPDs was then amplified by an SR570 low-noise amplifier (Stanford Research System) in high-bandwidth mode and eventually recorded by a DSOX3022T oscilloscope (Keysight Technologies).

The noise measurements of the devices were performed in the dark and the sample holder was shielded by a metal box to suppress the electromagnetic noises. The dark currents of the OPD at different external biases were amplified by a DLPCA-200 low-noise amplifier (FEMTO Messtechnik GmbH) and noise spectra were captured using a DSOX3022T oscilloscope operated with fast Fourier transform analysis.

**Thin film characterizations**

The thicknesses of the samples were collected with an Ambios XP-100 profilometer. AFM images in tapping mode were taken using a Bruker Multimode AFM 8 setup, using AFM tips with a force constant of 40 N/m and resonance frequency of c.a. 300 kHz.

Two-dimensional (2D) GIWAXS measurements were carried out at the PLS-II 5A beamline of the Pohang Accelerator Laboratory (PAL) in South Korea. The 2D images were taken by using a Mar 3450 CCD detector with a sample-to-detector distance of 414.415 mm at 11.57 keV (1.07 156 Å). The incidence angle ($α_i$) of the X-ray beam was selected between the critical angles of the thin film and substrate.

**Analytic formulas of RCE OPDs**

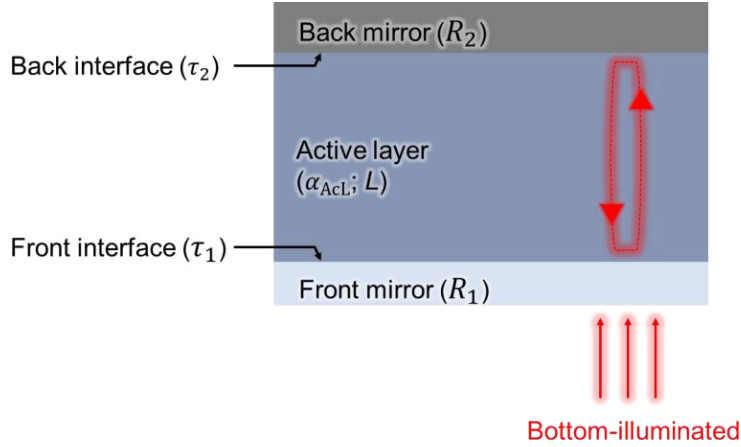

**Figure S2.** A schematic of bottom-illuminated RCE OPD with bottom contact acting as front mirror/interface and top contact acting as back mirror/interface.

Based on an analytical model for RCE photodetector proposed by Kishino *et al.*,[5] Tang *et al.* simplified and used the model to estimate photodetection parameters of a CT-state-based OPD with an active layer sandwiched between two mirrors.[6] Here, we adapt the model presented in ref. [6] to explain some observations found in our RCE OPDs.

The absorption (*A*) in the active layer of the cavity can be estimated as follows,[6]

$$A = \frac{e^{-\tau_1} + e^{-\tau_1} R_2 e^{-\alpha L}}{1 - 2\sqrt{R_1 R_2} e^{-\alpha L} \cos(\frac{4\pi L n}{\lambda}) + R_1 R_2 e^{-2\alpha L}} A_0, \quad (S1)$$

where *L* and *n* are the thickness and effective refractive index of the medium between two mirrors, respectively; $R_1$ and $R_2$ are the reflectance of the front and the back mirror, respectively; $\tau_1$ and $\tau_2$ are the optical depths of the front and the back mirror, respectively; $\alpha = (\tau_1 + \alpha_{AcL} L + \tau_2)/L$ is the effective absorption coefficient; $A_0 = (1 - R_1)(1 - e^{-\alpha_{AcL} L})$ is the fraction of light absorbed by a single pass through a non-enhanced active layer; and $\alpha_{AcL}$ is the absorption coefficient of the stand-alone active layer.

At the resonance condition, *i.e.* when $\lambda_m = \frac{2nL}{m}$ or $\cos(\frac{4\pi Ln}{\lambda}) = 1$, the absorption at the resonance wavelength now becomes,

$$A(\lambda_m) = \frac{e^{-\tau_1} + e^{-\tau_1}R_2 e^{-\alpha L}}{(1 - \sqrt{R_1 R_2}e^{-\alpha L})^2} A_0, \qquad (S2)$$

The full width at half maximum (FWHM) of $A(\lambda_m)$ peak can be found by solving for $\lambda$ at $A(\lambda_m)/2$, using **Equation (S1)** and **(S2)**,

$$\frac{A(\lambda_m)}{2} = \frac{e^{-\tau_1} + e^{-\tau_1}R_2 e^{-\alpha L}}{(1 - \sqrt{R_1 R_2}e^{-\alpha L})^2} A_0 = \frac{e^{-\tau_1} + e^{-\tau_1}R_2 e^{-\alpha L}}{1 - 2\sqrt{R_1 R_2}e^{-\alpha L}\cos(\frac{4\pi Ln}{\lambda}) + R_1 R_2 e^{-2\alpha L}} A_0. \qquad (S3)$$

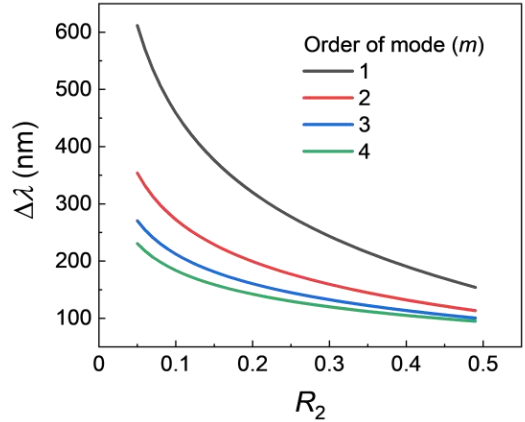

**Figure S3.** FWHM ($\Delta\lambda$) of resonant-enhanced absorption peak $A(\lambda_m)$ as a function of the reflectance of the front mirror $R_2$, using **Equation (S4)**.

From **Equation (S3)**, the FWHM value ($\Delta\lambda$) can be estimated as,

$$\Delta\lambda = \lambda_m \frac{1 - \sqrt{R_1 R_2}e^{-\alpha L}}{m\pi \sqrt[4]{R_1 R_2}e^{-0.5\alpha L}}. \qquad (S4)$$

In **Figure S3**, we show $\Delta\lambda$ as a function of the reflectance of the front mirror $R_2$ using **Equation (S4)**, assuming that $\alpha = 10^{-3}\ nm^{-1}$, $R_1 = 1$, and $m = 1, 2, 3$, and $4$ when $L = 80$, 280, 480, and 680 nm, respectively. Two clear trends that one can observe: (i) $\Delta\lambda$ decreases as $R_2$ increases, regardless of mode order $m$, and (ii) with the same value of $R_2$, an EQE peak with higher mode order is narrower. We can observe (ii) experimentally when comparing the EQE spectra near $\lambda = 1100$ nm of the $L_{BHJ} = 245$ nm ($m = 2$) and $L_{BHJ} = 655$ nm ($m = 4$) samples. The EQE at 1100 nm of the thicker sample appears as a sharp peak, which implies a narrow window of resonant-enhance EQE, while in the thinner sample, only a small bump can be observed (**Figures S11a** and **S11d**).

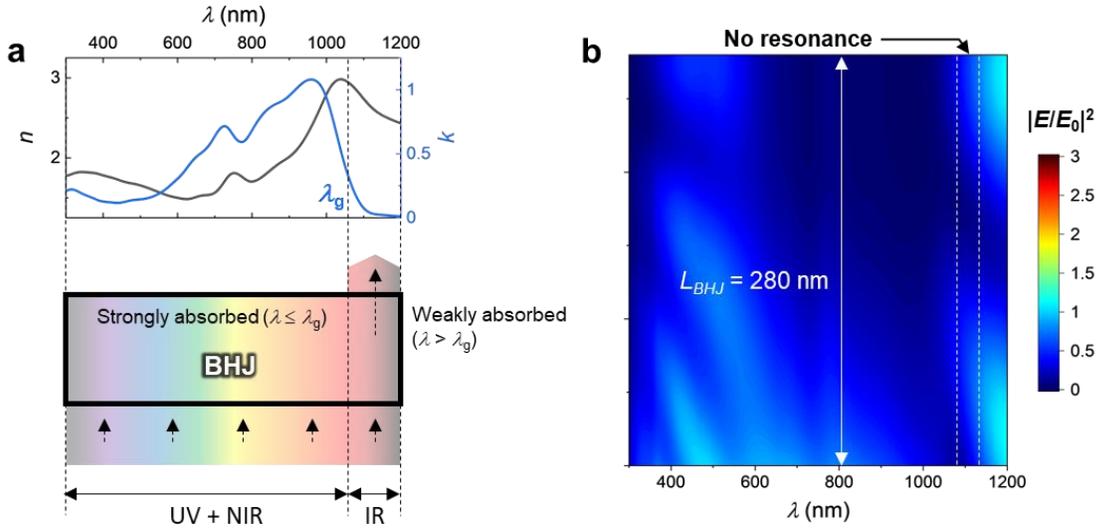

**Figure S4.** (a) *Top*: optical refractive index and extinction coefficient of PTB7-Th:COTIC-4F BHJ used for FDTD calculation. *Bottom*: Schematic of a stand-alone BHJ thin film. Photons with $\lambda < \lambda_g$ are strongly absorbed due to the high extinction coefficient of BHJ. Photons with $\lambda > \lambda_g$ are weakly absorbed and transmitted through the film. (b) Time-average intensity map of local electric field $|E/E_0|^2$ at the cross-section of a stand-alone BHJ film ($L_{BHJ} = 280$ nm).

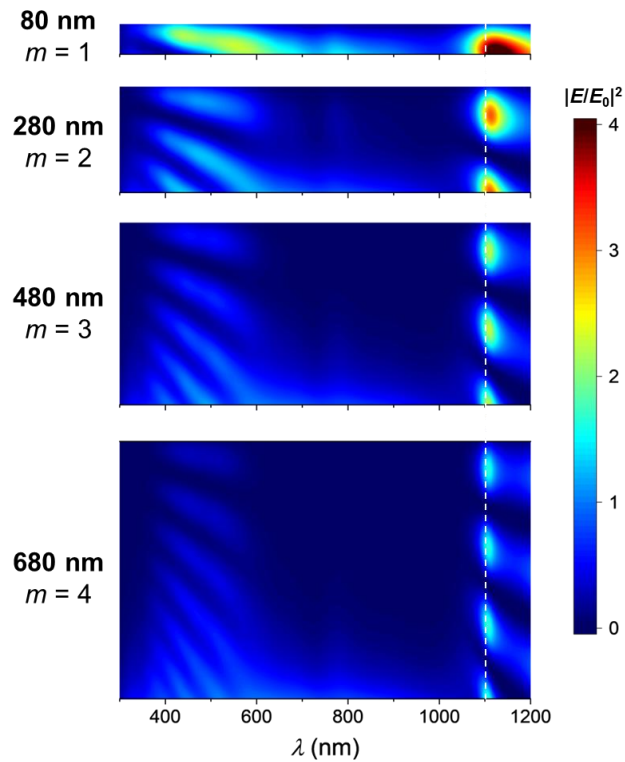

**Figure S5.** FDTD calculated time-average intensity map of the local electric field, at the cross-section of the OPD active layer with thicknesses of 80, 280, 480, and 680 nm, which supports resonance mode order $m$ = 1, 2, 3, and 4 at $\lambda_m$ = 1100 nm, respectively.

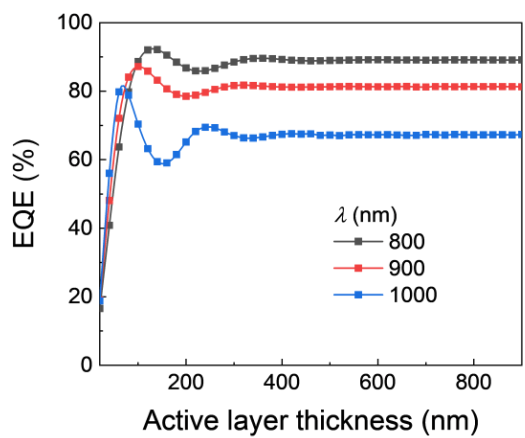

**Figure S6.** Calculated EQE versus active layer thickness at different $\lambda$, extract from **Figure 2b**.

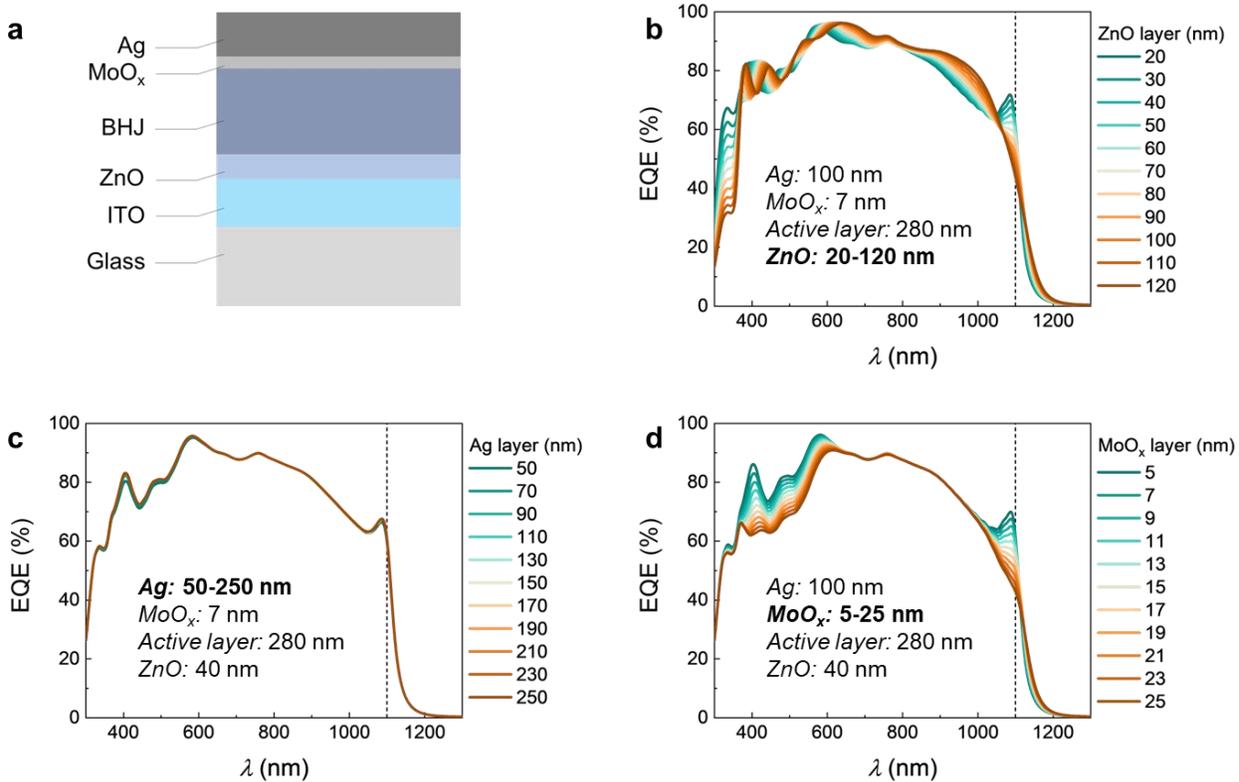

**Figure S7.** A schematic of the device structure used for FDTD calculation is illustrated in (a). Calculated EQE spectra as a function of (b) ZnO layer thickness, (c) Ag top electrode thickness, and (d) $MoO_x$ layer thickness.

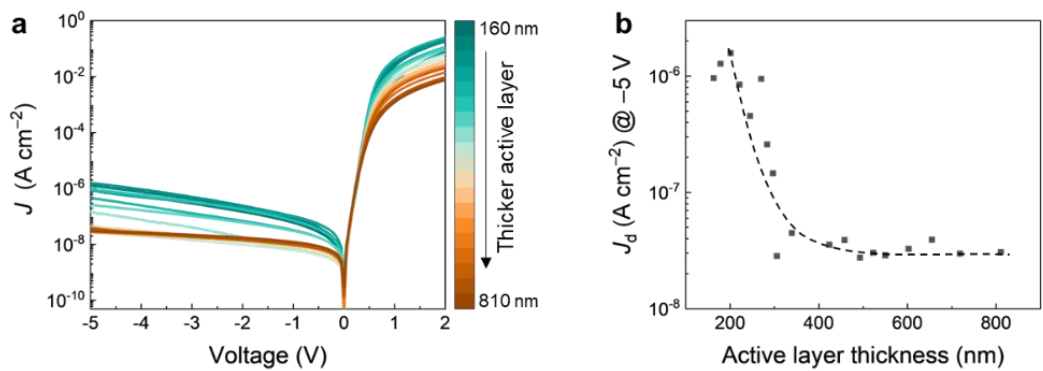

**Figure S8.** (a) Dark J-V characteristics and (b) $J_d$ at −5 V of PTB7-Th:COTIC-4F OPDs with different active layer thicknesses.

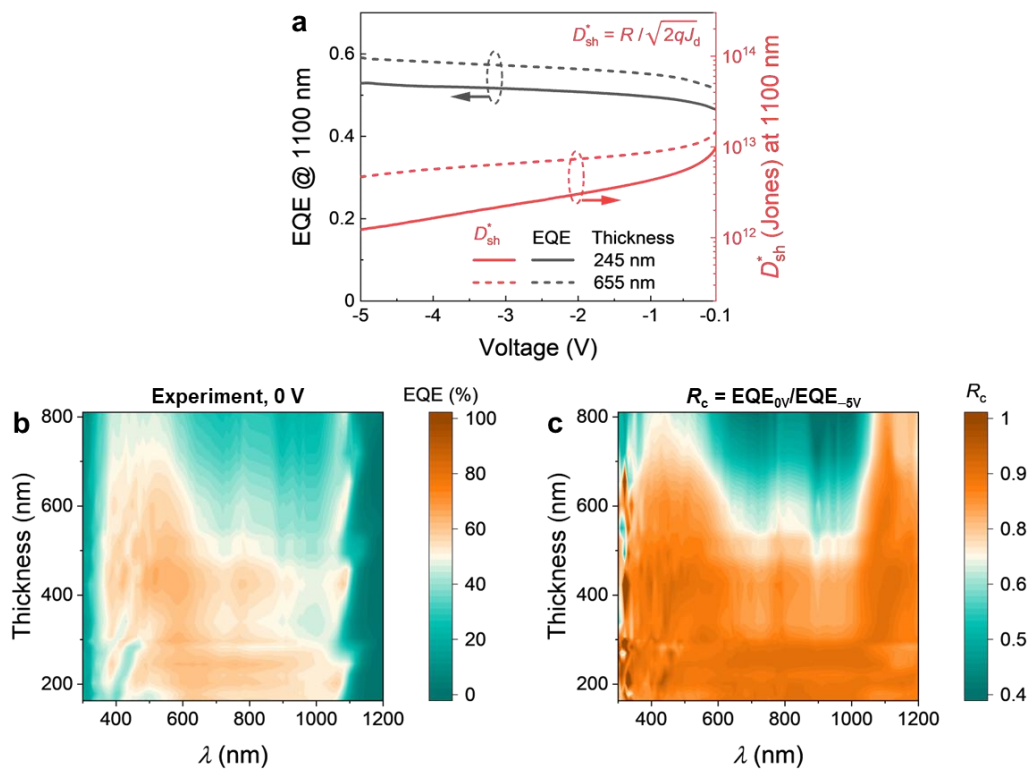

**Figure S9.** (a) Bias-dependent EQE and $D^*_{sh}$ (@ 1100 nm) of PTB7-Th:COTIC-4F OPDs with $L_{BHJ}$ = 245 and 655 nm. (b) Experimental EQE spectra (@0 V) and (c) $R_c$ = $EQE_{0V}/EQE_{-5V}$ of PTB7-Th:COTIC-4F OPDs with active layer thickness from 160 nm to

810 nm.

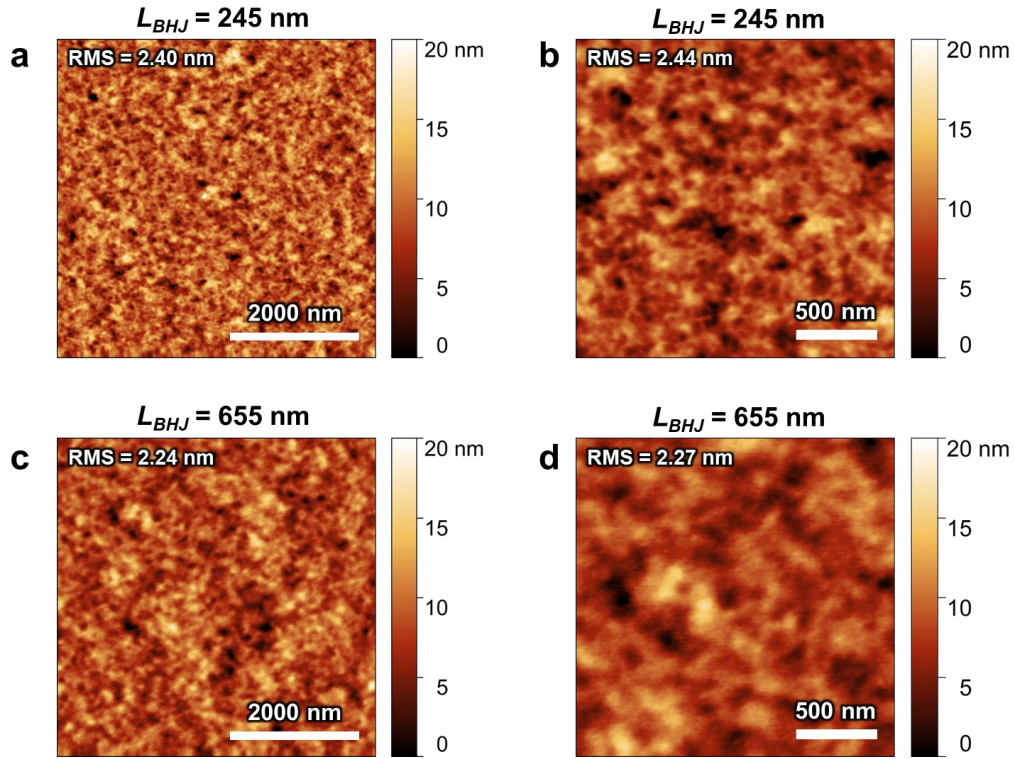

**Figure S10.** Tapping-mode atomic force microscope (AFM) images of the PTB7-Th:COTIC-4F BHJ films with (a-b) $L_{BHJ}$ = 245 nm and (c-d) $L_{BHJ}$ = 655 nm. Scan size of AFM images in (a) and (c) is 5 μm × 5 μm, and the scan size of AFM images in (b) and (d) is 2 μm × 2 μm.

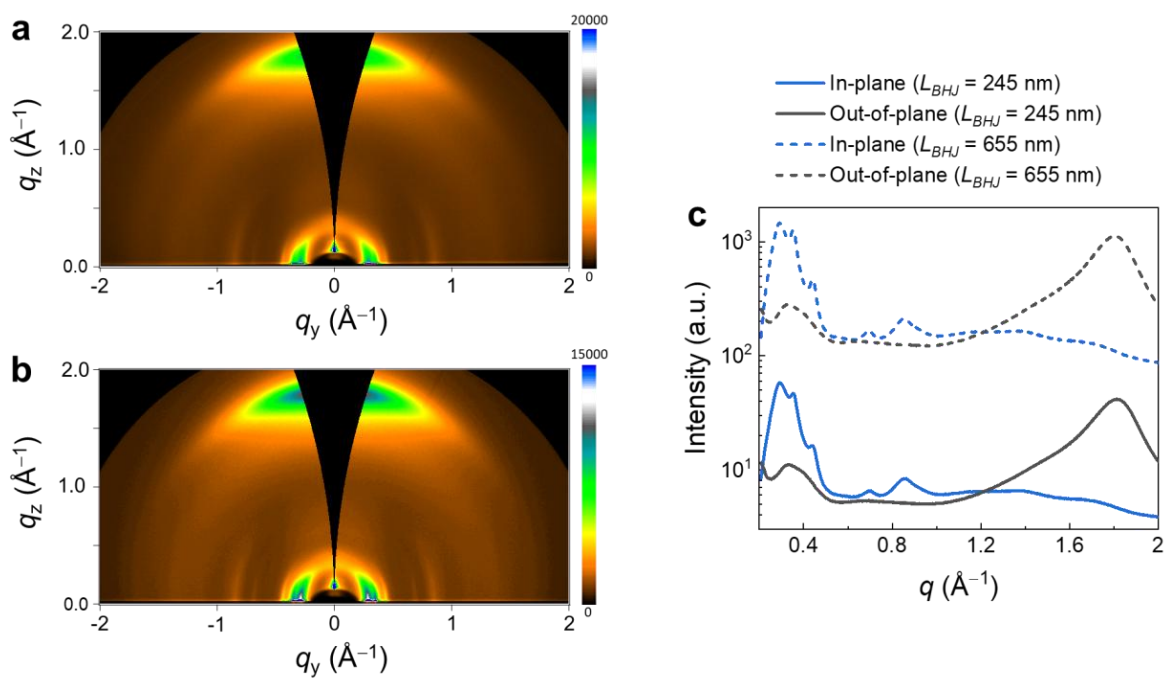

**Figure S11.** 2D GIWAXS images from the PTB7-Th:COTIC-4F blend films with two different thicknesses: (a) 245 nm and (b) 655 nm. (c) The out-of-plane and in-plane cuts are extracted from the 2D GIWAXS images.

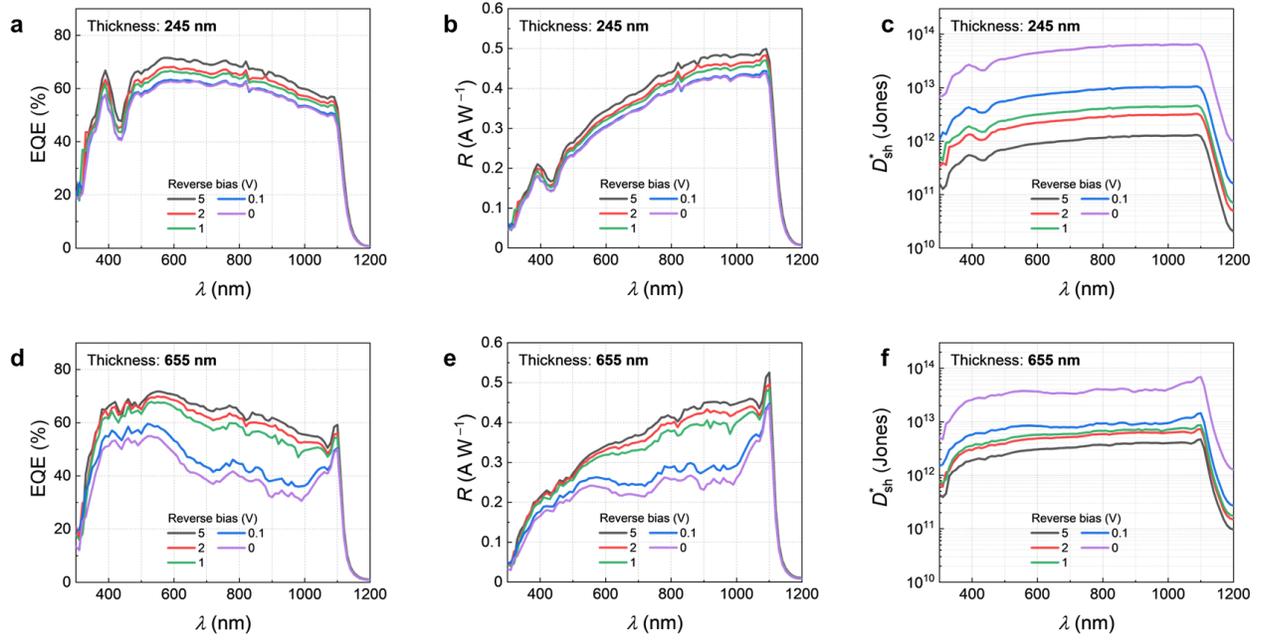

**Figure S12.** (a) EQE spectra, (b) $R$ spectra, and (c) $D^*_{sh}$ of PTB7-Th:COTIC-4F OPD with $L_{BHJ}$ = 245 nm at different reverse biases. (d) EQE spectra, (e) $R$ spectra, and (f) $D^*_{sh}$ of PTB7-Th:COTIC-4F OPD with $L_{BHJ}$ = 655 nm at different reverse biases.

We provide the shot-noise-limited specific detectivity at different reverse biases ($D^*_{sh} = \frac{R}{\sqrt{2qJ_d}}$), which can be derived directly from the dark current density (**Figure S12c and f**). It is worth noting that $D^*_{sh}$ is now widely used as a figure of merit for a simple comparison between OPD performances, rather than an estimation of the specific detectivity ($D^*$).[7] At a small bias of −0.1 V, both OPDs reach $D^*_{sh} > 10^{13}$ Jones at $\lambda$ = 1100 nm, and > $10^{12}$ Jones over a broad wavelength range of 300 – 1150 nm. At a large reverse bias of −5 V, thicker OPD benefits from much lower dark current density and the $D^*_{sh}$ at $\lambda$ = 1100 nm remains at 4.7 × $10^{12}$ Jones, which is about half an order of magnitude higher than the $D^*_{sh}$ of the thinner one (1.2 × $10^{12}$ Jones).

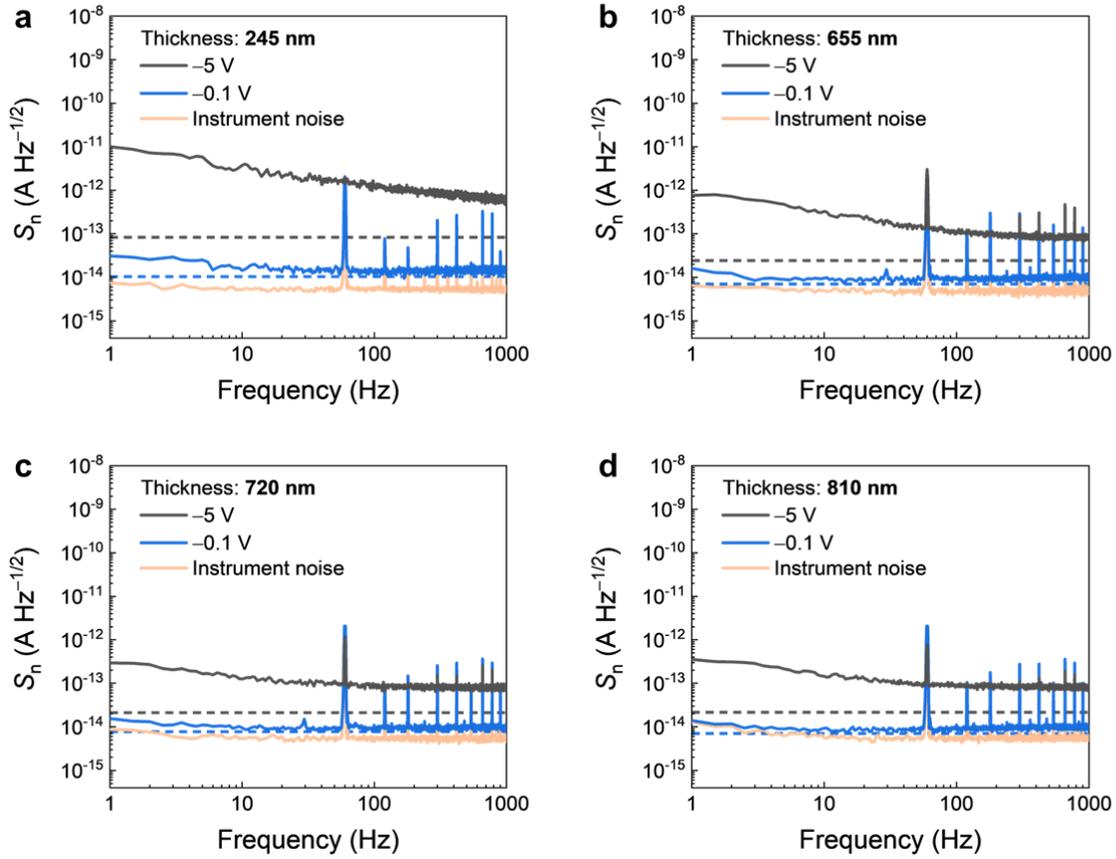

**Figure S13.** Measured noise current as a function of the frequencies at different applied biases for PTB7-Th:COTIC-4F OPDs with different active layer thicknesses. Blue and black dash lines represent the white noise levels $S_{wh} = \sqrt{S_{sh}^2 + S_{th}^2}$ at −0.1 V and −5 V as summarized in **Table S1**, respectively.

**Table S1.** Summary of noise current for PTB7-Th:COTIC-4F OPDs with different active layer thicknesses.

| Reverse bias (V) | Active layer thickness (nm) | $S_{n,thermal}$ (A Hz$^{-1/2}$) | $S_{n,shot}$ (A Hz$^{-1/2}$) | $S_{n,white}$ (A Hz$^{-1/2}$) | $S_{n,measure}$ @ 155 Hz (A Hz$^{-1/2}$) |
|---|---|---|---|---|---|
| 0.1 | 245 | 4.81E-15 | 9.12E-15 | 1.03E-14 | 1.44E-14 |
| | 655 | 2.00E-15 | 6.72E-15 | 7.01E-15 | 9.24E-14 |
| | 720 | 2.05E-15 | 7.46E-15 | 7.74E-15 | 9.62E-14 |
| | 810 | 2.20E-15 | 7.38E-15 | 7.70E-15 | 9.31E-14 |
| 5 | 245 | 4.81E-15 | 8.28E-14 | 8.29E-14 | 1.09E-12 |
| | 655 | 2.00E-15 | 2.43E-14 | 2.44E-14 | 1.02E-13 |
| | 720 | 2.05E-15 | 2.12E-14 | 2.13E-14 | 8.38E-14 |
| | 810 | 2.20E-15 | 2.15E-14 | 2.16E-14 | 8.69E-14 |
| Instrument background noise @ 155 Hz | | | | | 5.56E-15 |

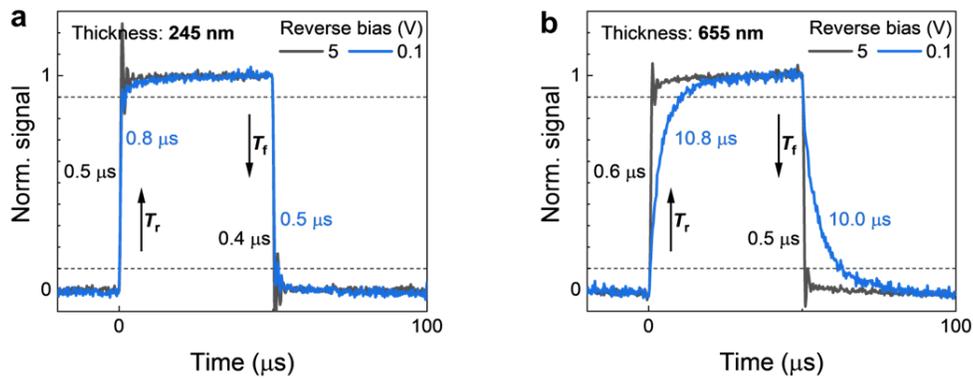

**Figure S14.** Transient responses under the illumination of a 1085 nm LED light source, and extracted rise/fall time as a function of reverse bias of PTB7-Th:COTIC-4F OPDs with (a)

$L_{BHJ}$ = 245 nm and (b) $L_{BHJ}$ = 655 nm.

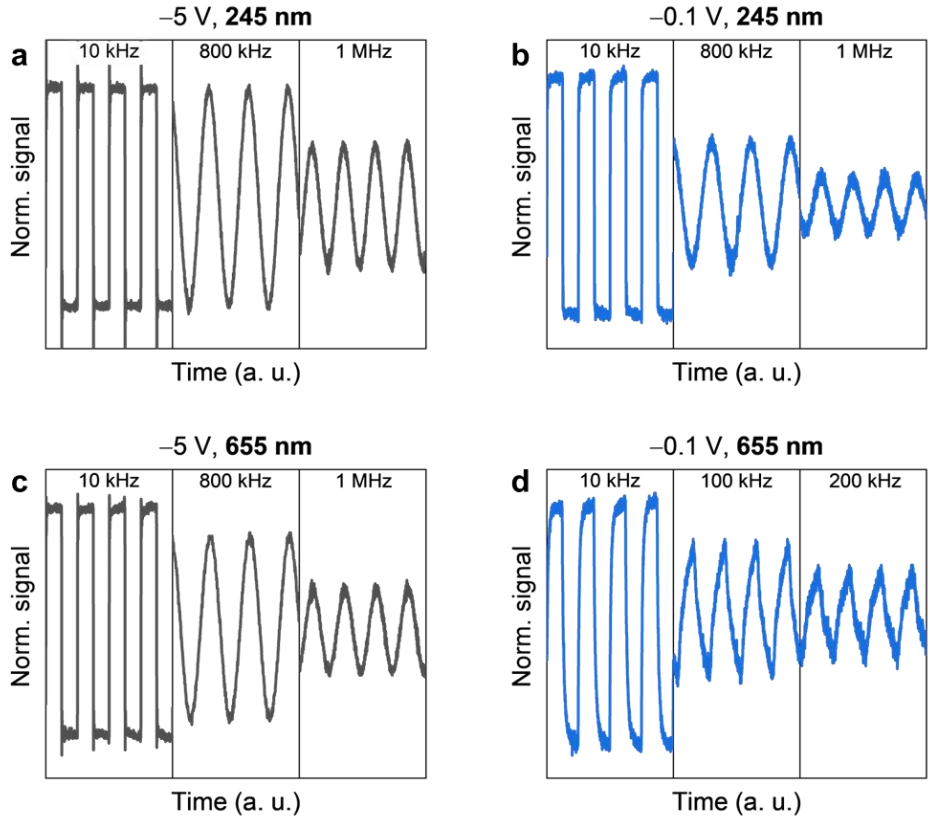

**Figure S15.** Photoresponse at different frequencies and reverse biases of PTB7-Th:COTIC-4F OPDs with $L_{BHJ}$ = 245 and 655 nm (under the illumination of a 1085 nm LED light source).

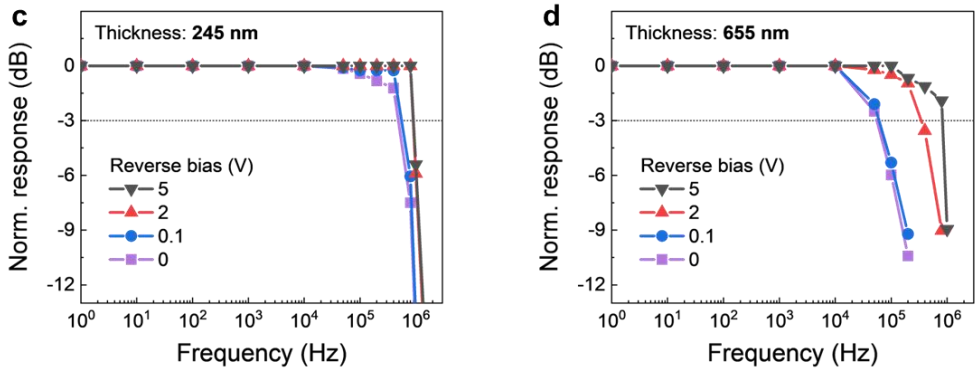

**Figure S16.** Normalized photoresponse as a function of input signal frequency and reverse bias of PTB7-Th:COTIC-4F OPDs with (a) $L_{BHJ} = 245$ nm and (b) $L_{BHJ} = 655$ nm.

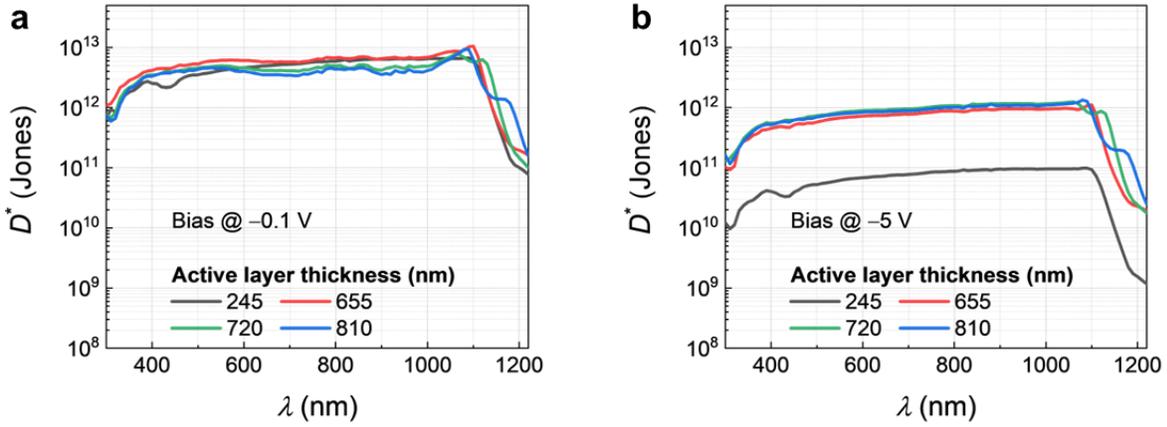

**Figure S17.** Specific detectivity $D^*$ spectra (derived from $S_n$) of PTB7-Th:COTIC-4F OPDs with different active layer thicknesses and under different reverse biases.

**Thickness optimization of PTB7-Th:IR6 OPDs**

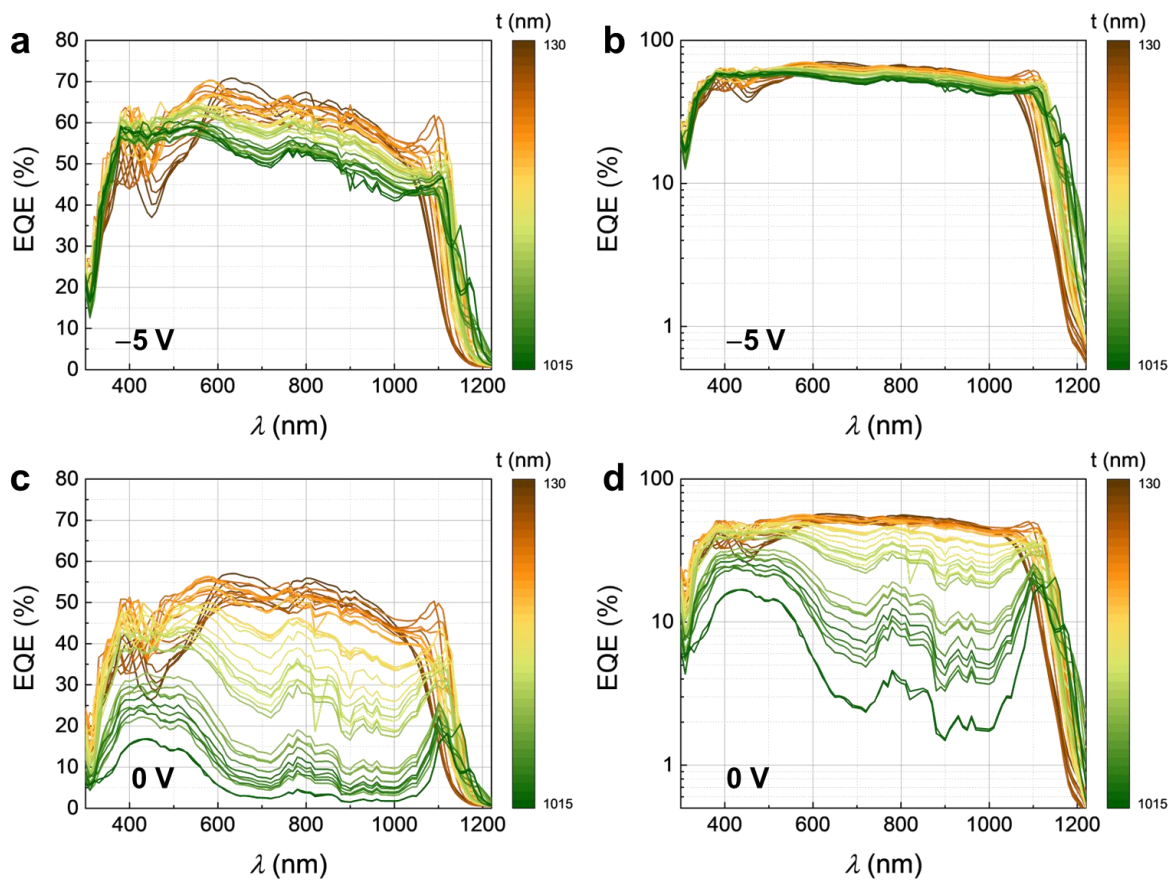

**Figure S18.** (a) Linear plot and (b) semilog plot of EQE spectra at −5 V of PTB7-Th:IR6 OPDs with different $L_{BHJ}$. (c) Linear plot and (d) semilog plot of EQE spectra at 0 V of PTB7-Th:IR6 OPDs with different $L_{BHJ}$.

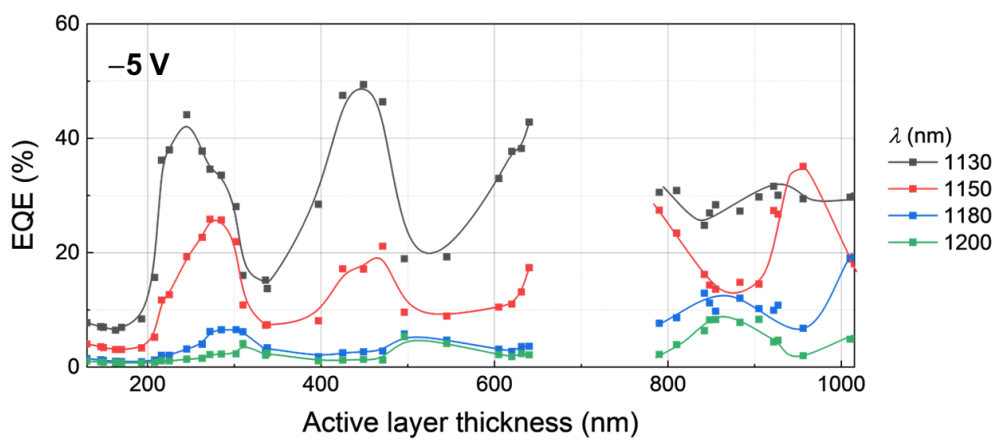

**Figure S19.** EQE (−5 V) of PTB7-Th:IR6 OPDs versus active layer thickness at different $\lambda$, extract from **Figure S18**.

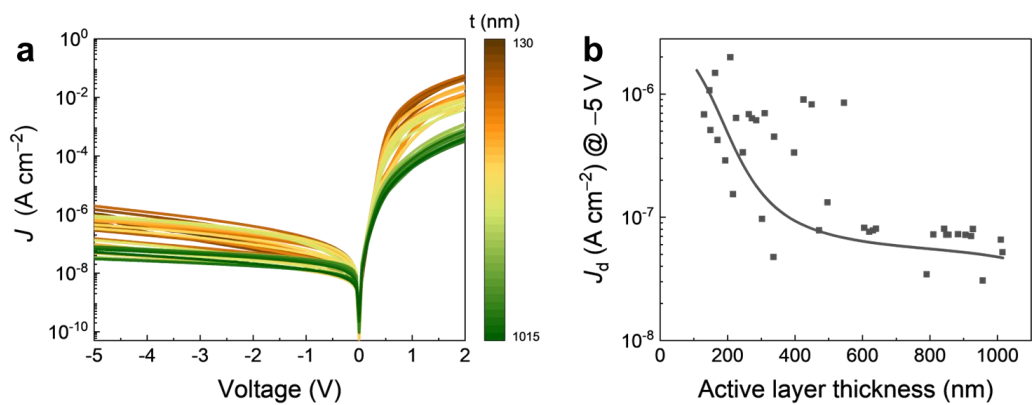

**Figure S20.** (a) Dark J-V characteristics and (b) $J_d$ at −5 V of PTB7-Th:IR6 OPDs with different active layer thicknesses.

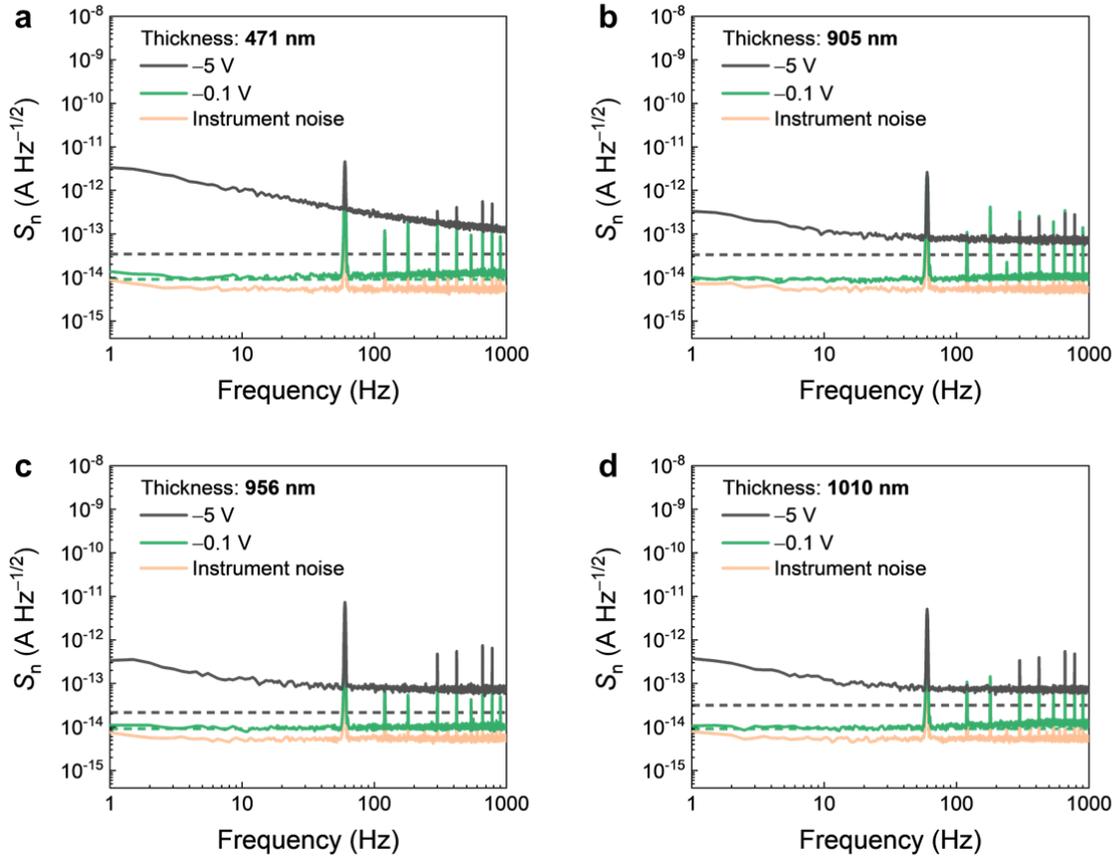

**Figure S21.** Measured noise current as a function of the frequencies at different applied biases for PTB7-Th:IR6 OPDs with different active layer thicknesses. Green and black dash lines represent the white noise levels $S_{wh} = \sqrt{S_{sh}^2 + S_{th}^2}$ at –0.1 V and –5 V, respectively.

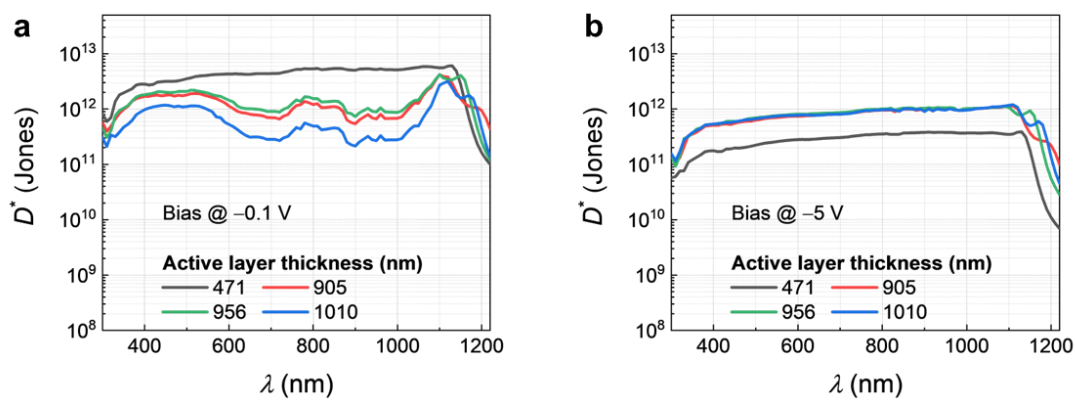

**Figure S22.** Specific detectivity $D^*$ spectra (derived from $S_n$) of PTB7-Th:IR6 OPDs with different active layer thicknesses and under different reverse biases.

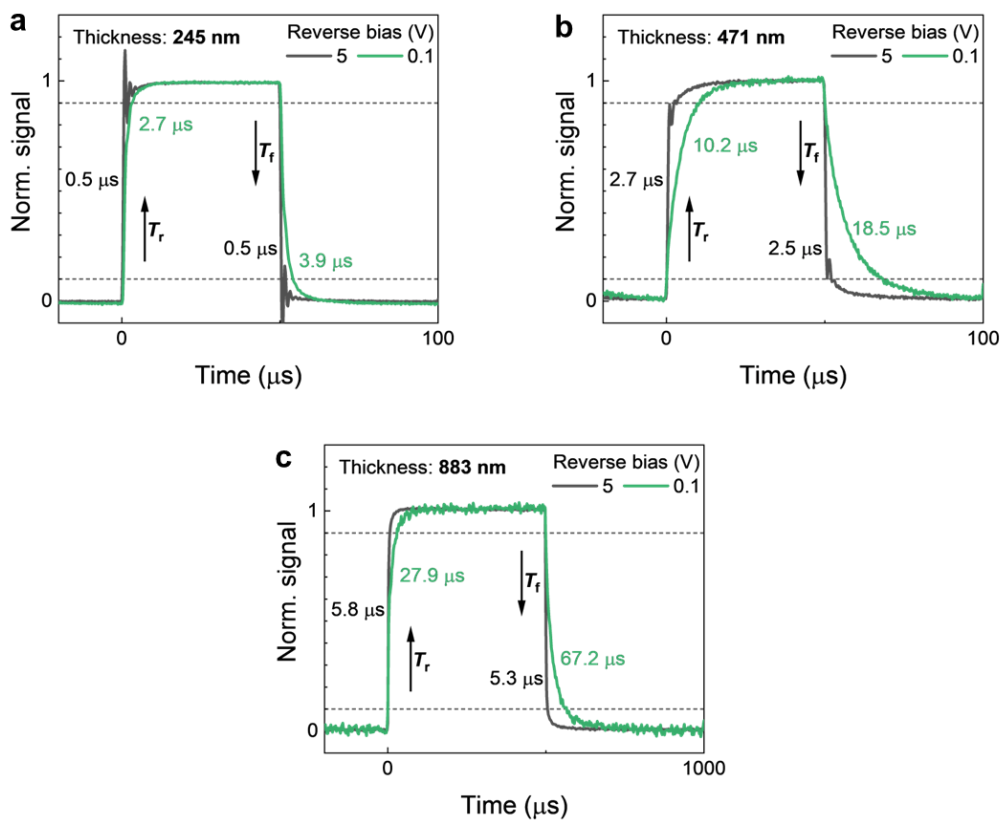

**Figure S23.** Transient responses under the illumination of a 1085 nm LED light source, and extracted rise/fall time as a function of reverse bias of PTB7-Th:IR6 OPDs with (a) $L_{BHJ}$ = 245 nm, (b) $L_{BHJ}$ = 471 nm, and (c) $L_{BHJ}$ = 883 nm.

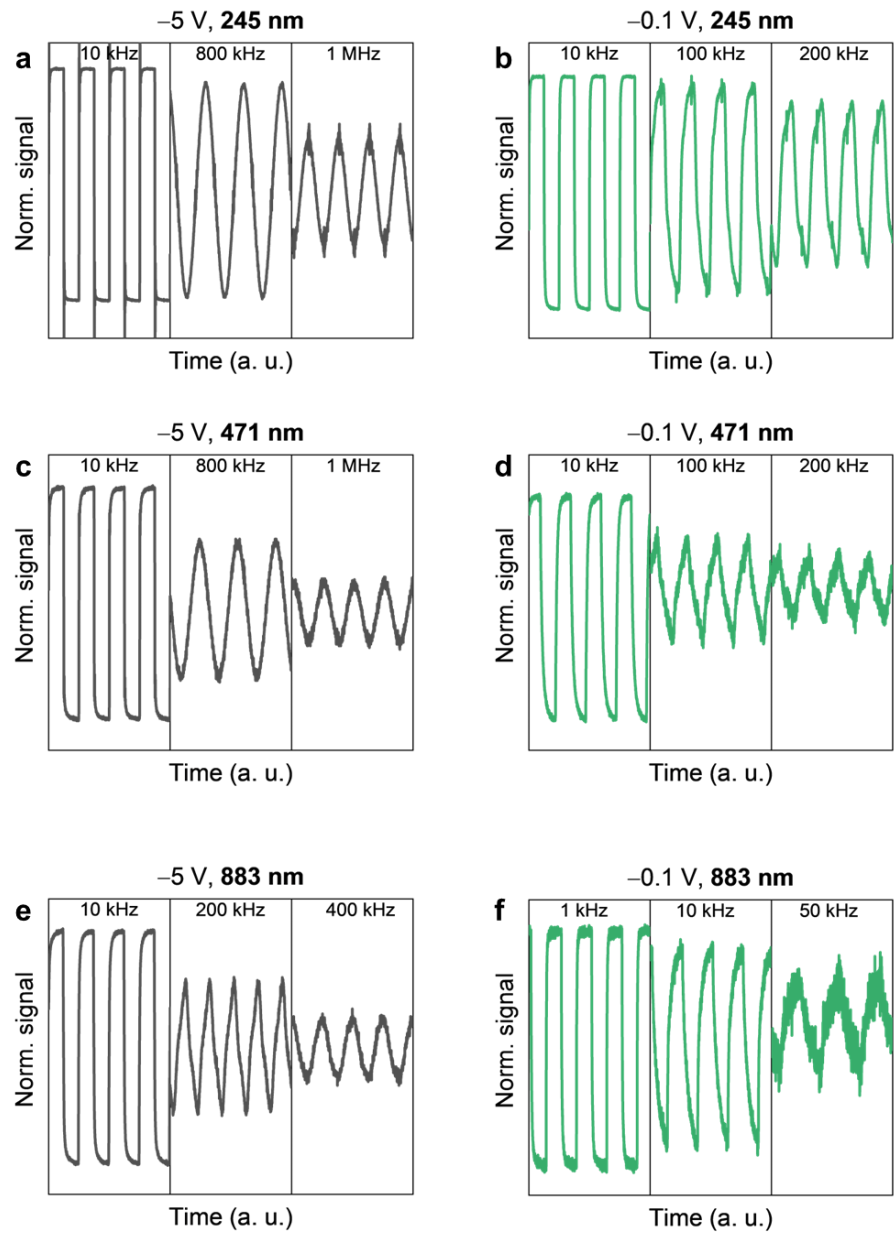

**Figure S24.** Photoresponse at different frequencies and reverse biases of PTB7-Th:IR6 OPDs with $L_{BHJ}$ = 245, 471, and 883 nm (under the illumination of a 1085 nm LED light source).

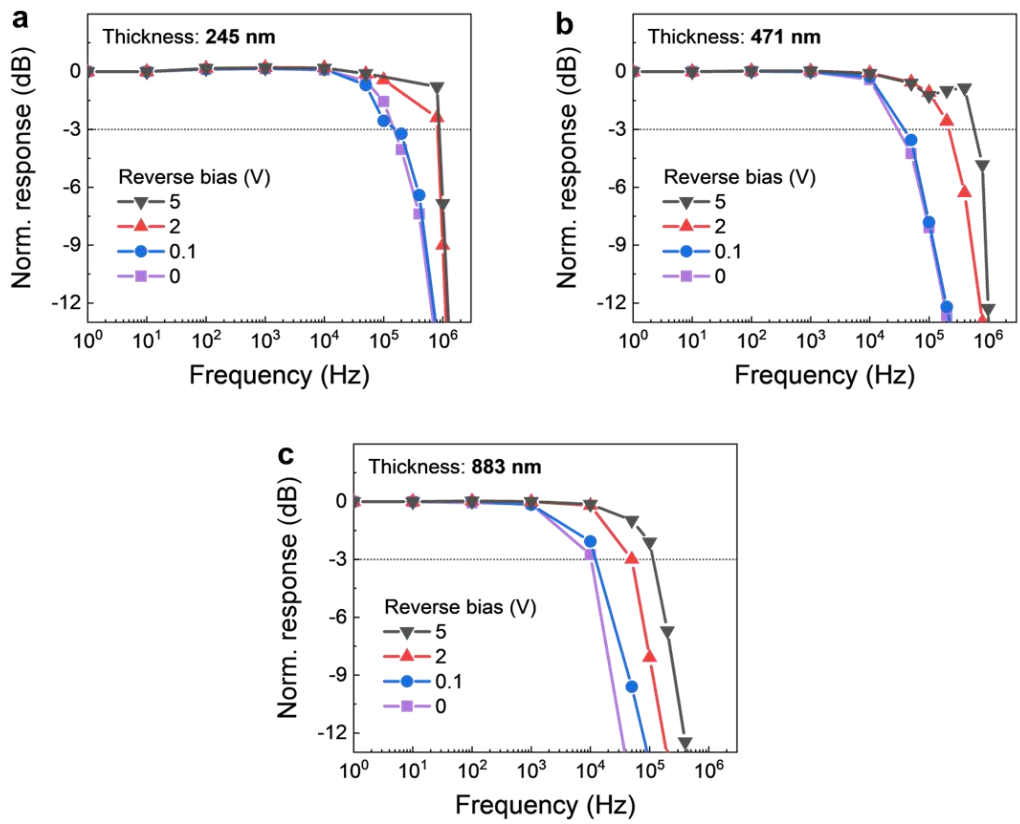

**Figure S25.** Normalized photoresponse as a function of input signal frequency and reverse bias of PTB7-Th:IR6 OPDs with (a) $L_{BHJ}$ = 245 nm, (b) $L_{BHJ}$ = 471 nm, and (c) $L_{BHJ}$ = 883 nm.

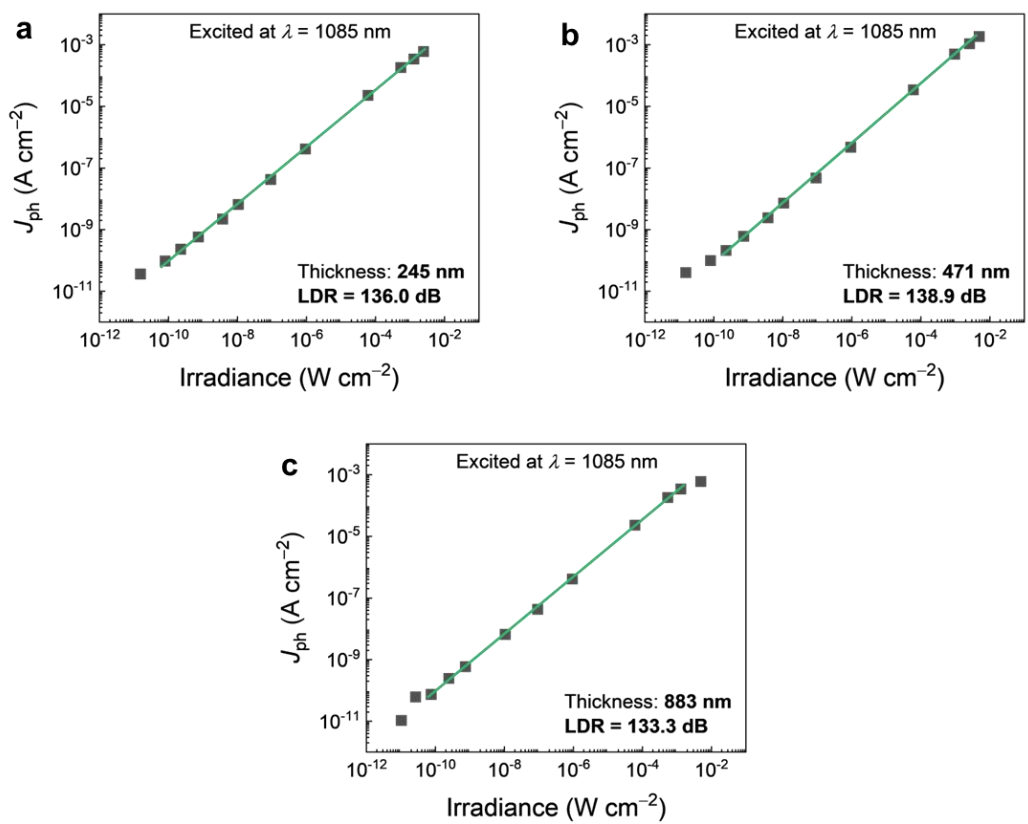

**Figure S26.** Linear dynamic range (LDR) at $\lambda = 1085$ nm of PTB7-Th:IR6 OPDs with (a) $L_{BHJ} = 245$ nm, (b) $L_{BHJ} = 471$ nm, and (c) $L_{BHJ} = 883$ nm. Applied bias is –0.1 V.

**Table S2.** NFA-based OPDs presented in **Figure 6**.

| NFA-based BHJ | EQE (%) | | | $D^*$ or $D^*_{sh}$ (Jones) | | | Broad-band? | Ref. |
|---|---|---|---|---|---|---|---|---|
| | Value | Bias | $\lambda$ | Value | Bias | $\lambda$ | | |
| PTB7-Th:COTIC-4Cl:PC$_{71}$BM | ~42 | 0 | 1060 | 6.00E+12 | 0.1 | 1060 | Yes | [8] |
| PTB7-Th:NTQ | 30 | 0.1 | 1000 | 3.72E+12 | 0.1 | 1000 | Yes | [9] |
| PTB7-Th:COTIC-4F:Y6 | 48 | 0.1 | 1060 | 2.10E+13 | 0.1 | 1060 | Yes | [10-11] |
| PD-940 ink | 74 | 4 | 1000 | 6.60E+12 | 4.0 | 1000 | Yes | [12] |
| PD-940 ink | 27.3 | 0 | 1030 | 1.91E+12 | 0.0 | 1030 | Yes | [13] |
| | 70.7 | 4 | 1030 | 2.74E+12 | 4.0 | 1030 | Yes | |
| | 77.2 | 10 | 1030 | 4.6E+12 | 10.0 | 1030 | Yes | |
| PTB7-Th:COTIC-4F | 18 | 0 | 1096 | 6.59E+12 | 0.0 | 1096 | No | [14] |
| PTB7-Th:SiOTIC-4F | 20 | 0 | 1026 | 1.23E+13 | 0.0 | 1026 | No | |
| PTB7-Th:IEICO-4F | 18 | 0 | 980 | 7.02E+12 | 0.0 | 980 | No | |
| PD-940 ink | 53 | 8 | 1080 | 2.34E+12 | 8.0 | 1080 | No | [15] |
| PTB7-Th:COTIC- | 58.6 | 0 | 1060 | | | | Yes | **This** |
| 4F | 58.9 | 0 | 1080 | | | | Yes | **work** |
| | 50 | 0 | 1100 | | | | Yes | |
| | 30.8 | 0 | 1120 | | | | Yes | |
| | 25.9 | 0 | 1130 | | | | Yes | |
| | 16.3 | 0 | 1140 | | | | Yes | |
| | 7.03 | 0 | 1160 | | | | Yes | |
| | 5.4 | 0 | 1180 | | | | Yes | |
| | 1.96 | 0 | 1200 | | | | Yes | |
| | 67.3 | 5 | 1060 | | | | Yes | |
| | 67.2 | 5 | 1080 | | | | Yes | |
| | 59.2 | 5 | 1100 | | | | Yes | |
| | 37.4 | 5 | 1120 | | | | Yes | |
| | 33.6 | 5 | 1130 | | | | Yes | |
| | 20.2 | 5 | 1140 | | | | Yes | |
| | 8.44 | 5 | 1160 | | | | Yes | |
| PTB7-Th:COTIC-4F | 7 | 5 | 1180 | | | | Yes | **This work** |
| | 2.67 | 5 | 1200 | | | | Yes | |

|  |  |  |  | 1.40E+13 | 0.1 | 1100 | Yes |  |
|  |  |  |  | 6.86E+12 | 0.1 | 1130 | Yes |  |
|  |  |  |  | 2.00E+12 | 0.1 | 1150 | Yes |  |
|  |  |  |  | 1.50E+12 | 0.1 | 1180 | Yes |  |
|  |  |  |  | 1.1E+13 ($S_n$ derived) | 0.1 | 1100 | Yes |  |
|  |  |  |  | 5.3E+12 ($S_n$ derived) | 0.1 | 1130 | Yes |  |
|  |  |  |  | 1.5E+12 ($S_n$ derived) | 0.1 | 1150 | Yes |  |
|  |  |  |  | 1.2E+12 ($S_n$ derived) | 0.1 | 1180 | Yes |  |
| PTB7-Th:IR6 | 48.9 | 0 | 1100 |  |  |  | Yes | **This work** |
|  | 48.1 | 0 | 1110 |  |  |  | Yes |  |
|  | 41.4 | 0 | 1120 |  |  |  | Yes |  |
|  | 37.8 | 0 | 1130 |  |  |  | Yes |  |
|  | 20.5 | 0 | 1150 |  |  |  | Yes |  |
| PTB7-Th:IR6 | 16.6 | 0 | 1160 |  |  |  | Yes | **This work** |
|  | 9.7 | 0 | 1170 |  |  |  | Yes |  |
|  | 8.2 | 0 | 1180 |  |  |  | Yes |  |
| | 5.9 | 0 | 1190 | | | | Yes | |
| | 4.6 | 0 | 1200 | | | | Yes | |
| | 3.3 | 0 | 1210 | | | | Yes | |
| | 2.3 | 0 | 1220 | | | | Yes | |
| | 60.1 | 5 | 1100 | | | | Yes | |
| | 59.6 | 5 | 1110 | | | | Yes | |
| | 53.3 | 5 | 1120 | | | | Yes | |
| | 49.4 | 5 | 1130 | | | | Yes | |
| | 35 | 5 | 1150 | | | | Yes | |
| | 28.3 | 5 | 1160 | | | | Yes | |
| | 22.4 | 5 | 1170 | | | | Yes | |
| | 19.1 | 5 | 1180 | | | | Yes | |
| | 10.7 | 5 | 1190 | | | | Yes | |
| | 8.34 | 5 | 1200 | | | | Yes | |
| PTB7-Th:IR6 | 6.02 | 5 | 1210 | | | | Yes | **This work** |
| | 3.74 | 5 | 1220 | | | | Yes | |
| | | | | 6.1E+12 | 0.1 | 1130 | Yes | |

| | | | | | | | |
|---|---|---|---|---|---|---|---|
| | | | | ($S_n$ derived) | | | |
| | | | | 4.1E+12 ($S_n$ derived) | 0.1 | 1150 | Yes |
| | | | | 1.5E+12 ($S_n$ derived) | 0.1 | 1180 | Yes |
| | | | | 9.4E+11 ($S_n$ derived) | 0.1 | 1200 | Yes |

## Supplementary References

[1] J. Lee, S. J. Ko, M. Seifrid, H. Lee, B. R. Luginbuhl, A. Karki, M. Ford, K. Rosenthal, K. Cho, T. Q. Nguyen, *Advanced Energy Materials* **2018**, 8, 1801212.

[2] FDTD Solutions, https://www.lumerical.com/, accessed.

[3] N. Schopp, V. V. Brus, J. Lee, G. C. Bazan, T. Q. Nguyen, *Advanced Energy Materials* **2021**, 11, 2002760.

[4] Solar cell methodology - Ansys Optics - FDTD Lumerical, https://optics.ansys.com/hc/en-us/articles/360042165634, accessed: 2022.

[5] K. Kishino, M. S. Unlu, J.-I. Chyi, J. Reed, L. Arsenault, H. Morkoc, *IEEE Journal of Quantum Electronics* **1991**, 27, 2025.

[6] Z. Tang, Z. Ma, A. Sánchez‐Díaz, S. Ullbrich, Y. Liu, B. Siegmund, A. Mischok, K. Leo, M. Campoy‐Quiles, W. Li, *Adv. Mater.* **2017**, 29, 1702184.

[7] J. Liu, M. Gao, J. Kim, Z. Zhou, D. S. Chung, H. Yin, L. Ye, *Materials Today* **2021**.

[8] Z. Zhong, F. Peng, L. Ying, G. Yu, F. Huang, Y. Cao, *Science China Materials* **2021**, 64, 2430.

[9] S. Deng, L. Zhang, J. Zheng, J. Li, S. Lei, Z. Wu, D. Yang, D. Ma, J. Chen, *Adv. Opt. Mater.* **2022**, 2200371.

[10] Y. Song, Z. Zhong, P. He, G. Yu, Q. Xue, L. Lan, F. Huang, *Adv. Mater.* **2022**, 2201827.

[11] C. Xu, P. Liu, C. Feng, Z. He, Y. Cao, *Journal of Materials Chemistry C* **2022**, 10, 5787.

[12] J. L. Wu, L. H. Lai, Y. T. Hsiao, K. W. Tsai, C. M. Yang, Z. W. Sun, J. C. Hsieh, Y. M. Chang, *Adv. Opt. Mater.* **2022**, 10, 2101723.

[13] L.-H. Lai, C.-C. Hsieh, J.-L. Wu, Y.-M. Chang, *ACS Applied Electronic Materials* **2021**, 4, 168.

[14] Q. Liu, S. Zeiske, X. Jiang, D. Desta, S. Mertens, S. Gielen, R. Shanivarasanthe, H.-G. Boyen, A. Armin, K. Vandewal, *Nat. Commun.* **2022**, 13, 1.

[15] K.-W. Tsai, G. Madhaiyan, L.-H. Lai, Y.-T. Hsiao, J.-L. Wu, C.-Y. Liao, C.-H. Hou, J.-J. Shyue, Y.-M. Chang, *ACS Appl. Mater. Interfaces* **2022**, 14, 38004.